\documentclass[aps, prd, twocolumn, superscriptaddress, preprintnumbers, nofootinbib,longbibliography,10pt]{revtex4-1}

\usepackage{amsmath}	
\usepackage{amsthm}		
\usepackage{amssymb}	
\usepackage{datetime}
\usepackage{graphicx}
\usepackage{color}
\usepackage{verbatim}
\usepackage{bm}
\usepackage{slashed}

\usepackage[colorlinks=true, citecolor=midblue, linkcolor=midblue, urlcolor=midblue]{hyperref}

\usepackage{setspace}

\usepackage{tikz,xcolor}
\definecolor{lime}{HTML}{A6CE39}
\DeclareRobustCommand{\orcidicon}{
	\begin{tikzpicture}
	\draw[lime, fill=lime] (0,0) 
	circle [radius=0.16] 
	node[white] {{\fontfamily{qag}\selectfont \tiny ID}};
	\draw[white, fill=white] (-0.0625,0.095) 
	circle [radius=0.007];
	\end{tikzpicture}
	\hspace{-2mm}
}
\foreach \x in {A, ..., Z}{
	\expandafter\xdef\csname orcid\x\endcsname{\noexpand\href{https://orcid.org/\csname orcidauthor\x\endcsname}{\noexpand\orcidicon}}
}

\settimeformat{ampmtime}

\definecolor{grey}{rgb}{0.4,0.4,0.4}
\definecolor{dullmagenta}{rgb}{0.4,0,0.4}
\definecolor{darkblue}{rgb}{0,0,0.4}
\definecolor{midblue}{rgb}{0,0,0.5}
\definecolor{midred}{rgb}{0.5,0,0}
\definecolor{orange}{rgb}{1,0.5,0}
\definecolor{lightbrown}{rgb}{0.75,0.5,0.25}
\definecolor{tan}{cmyk}{0.14,0.42,0.56,0}
\definecolor{djunglegreen}{cmyk}{0.99,0,0.52,0}
\definecolor{lightgreen}{rgb}{0,1,0}
\definecolor{olivegreen}{cmyk}{0.64,0,0.95,0.40}
\definecolor{midgreen}{rgb}{0.0,0.675,0.0}
\definecolor{darkgreen}{rgb}{0,0.5,0}

\settimeformat{ampmtime}

\usepackage[notrig]{physics}

\newcommand{\FirstAffiliation}{\affiliation{
	Arnold Sommerfeld Center,
	Ludwig-Maximilians-Universit{\"a}t,
	Theresienstra{\ss}e 37,
	80333 M{\"u}nchen,
	Germany}}
 
\newcommand{\SecondAffiliation}{\affiliation{
	Max-Planck-Institut f{\"u}r Physik,
	Boltzmannstrasse 8,
	85748 Garching ,
	Germany}}

 \newcommand{\ThirdAffiliation}{\affiliation{    
Sydney Consortium for Particle Physics and Cosmology, 
School of Physics, The University of Sydney, NSW 2006, Australia}}

\usetikzlibrary {arrows.meta}

\begin{document}
\date{\formatdate{\day}{\month}{\year}, \currenttime}

\title{Similarities in the evaporation of saturated solitons and black holes}

	\author{Giacomo Contri\orcidA{}}
	\email{contri@mpp.mpg.de}
	\SecondAffiliation
\FirstAffiliation

 \author{Gia Dvali}
 	\SecondAffiliation
	\FirstAffiliation

	\author{Otari Sakhelashvili}
        \thanks{The current address is 1}
        \SecondAffiliation
	\ThirdAffiliation

	\date{\small\today} 
 
	\begin{abstract} 	
		\noindent  
        It has been suggested some time ago that many black hole properties are not specific to gravity, but rather are shared by a large class of objects, the so-called saturons, that saturate the quantum field theoretic upper bound on microstate degeneracy. By now, various aspects of this universality have been understood and demonstrated in a number of explicit examples. In the present paper, we add one more brick to the building by showing that the decay of a simple two-dimensional saturated soliton copies some key aspects of the black hole decay as well as of the information retrieval.  In particular, we study the evaporation process of a classically-stable vacuum bubble of a spontaneously broken $SU(N)$-symmetry, coupled to massless fermions. We show that the decay rate as well as the characteristic energy of the emitted quanta are given by the inverse size of the object, in striking similarity with the Hawking evaporation of a black hole. The time-scale of information retrieval also matches the one previously suggested for a black hole by Page. We give the semiclassical derivation of the phenomenon as well as its fully quantum resolution as a decaying coherent state of Goldstone bosons. The universal nature of the effect and its microscopic understanding support the analogous quantum portrait of a black hole as a saturated coherent state of gravitons.

	\end{abstract}
 \maketitle 
 
	\section{Introduction}
	Black holes have long been considered as objects in their own category, characterized by certain unique properties. Perhaps, the feature that has been considered to be the most mysterious is the  Bekenstein-Hawking entropy, which follows an area law in Planck units \cite{Bekenstein:1973ur},
    \begin{equation} \label{BHent}
		S \sim A \, M_P^{d-2} \,,
	\end{equation}
    where $d$ is the dimensionality of the space-time and $M_P$ the corresponding Planck mass.  
    This entropy also saturates the Bekenstein Bound \cite{Bekenstein:1980jp} $S\leq 2\pi ER$, where $E$ is the energy and $R$ is the radius of the object. 
    
    The black hole exhibits other striking features. 
    In the semiclassical description it is characterized by an information horizon as well as by the thermal-like Hawking evaporation rate \cite{Hawking:1975vcx}, 
     \begin{equation}\label{eq:thermal_rate}
    \Gamma \sim \frac{1}{R} \,.
\end{equation}
 Both the rate as well as the characteristic energy of emitted quanta are given by the inverse size of the system. 

  Another mysterious property of a black hole is a macroscopically-long time-scale of information retrieval. Namely, despite having a maximal information storage capacity, a semiclassical black hole radiates energy but no (or negligible) information. The duration of the validity of the semiclassical regime, which ignores the quantum backreaction, is a separate important question entangled 
  with the question of information retrieval. 
   This shall be discussed separately. 
   However, there is little doubt that, within the validity of the semiclassical treatment, the information carried by radiation is unreadable. This is puzzling. 
   
 One may be tempted to attribute the above features to peculiarities of quantum gravity, however this turned out not to be the case. 
 
 First, in \cite{Dvali:2018xpy} it was shown that the feature of the delayed information-retrieval is a universal property 
 of all systems with enhanced capacity of information storage, regardless of their origin. 
 All such systems at the initial stages radiate energy but maintain the load of information internally for a macroscopically long time-scale. However, the universality of black hole features does not stop here. 

The unifying framework was proposed in \cite{Dvali:2020wqi,Dvali:2019jjw,Dvali:2019ulr}, where it was argued that none of the above-discussed features are unique to black holes. Rather, they are common for a universality class of objects, the so-called ``saturons", that saturate the quantum field theoretic (QFT) bound on the microstate degeneracy.  In \cite{Dvali:2020wqi,Dvali:2019jjw,Dvali:2019ulr} the bound has been formulated in three distinct forms, which emphasize different physical meanings of the saturation effect. 
 
 First, the bound says that in any QFT the maximal microstate entropy attainable by a localized object of radius $R$ is given by   
    \begin{equation}\label{coupling_bound}
		S_{\rm max} \, = \, \frac{1}{\tilde{\alpha}} \,,
	\end{equation}
where $\tilde{\alpha}$ is a dimensionless quantum coupling of the interaction responsible for the existence of the object. In the above expression, the running coupling $\tilde{\alpha}$ has to be evaluated at the scale $R$.
    
	Moreover, $\tilde{\alpha}$ also sets the upper bound on the occupation number of quanta $N$ in the bound state by the self-sustainability condition $\tilde{\alpha} N = 1$.
 Thus, for a saturated system the microstate entropy is bounded by the number of constituents 
 \cite{Dvali:2019jjw}, 
   \begin{equation}\label{NBound}
		S_{\rm max} \, = \, N  \,.
	\end{equation}
    
    The alternative (and perhaps a more striking) presentation of the same bound 
 is that the maximal entropy is given by the area 
  \begin{equation} \label{Areaf}
		S_{\rm max} \, = \, A \, f_P^{d-2} \,, 
	\end{equation}
 where $f_P$ is the scale of spontaneous breaking of the Poincaré symmetry by the object. 
 
 The connection between (\ref{Areaf})  and (\ref{coupling_bound}) becomes very clear by noticing that the scale $f_P$ is the scale (``decay constant") of the Poincaré Goldstone. 
 Correspondingly, the dimensionless coupling of this Goldstone, evaluated at the momentum scale $1/R$ is (\ref{coupling_bound}). In other words, at the point when a system reaches the maximal microstate degeneracy, the coupling $\tilde{\alpha}$ of the interaction responsible for the existence of the object matches the coupling of the Goldstone mode of the spontaneously-broken Poincaré symmetry. 
 
 It has been shown that going beyond the bounds (\ref{coupling_bound}), (\ref{NBound})
 and (\ref{Areaf}) invalidates the QFT description, in particular, by the
 breakdown of the loop-expansion in $\tilde{\alpha}$
 \cite{Dvali:2019jjw,Dvali:2019ulr},
 and by the violation of unitarity in the scattering processes
 \cite{Dvali:2020wqi}.

   Therefore, the phenomenon of saturation gives a transparent meaning to the area-form of the entropy: it manifests the universal feature of the (inverse) Goldstone coupling which at wavelength $R$ in an arbitrary number of space-time dimensions scales as the corresponding area $A \sim R^{d-2}$.

  Once we understand the above, it becomes immediately clear that the black hole entropy (\ref{BHent}) 
  is a particular manifestation of the more general expressions (\ref{Areaf}) and (\ref{coupling_bound})
 \cite{Dvali:2020wqi,Dvali:2019jjw,Dvali:2019ulr}.  
 Indeed, for equating (\ref{BHent}) and (\ref{Areaf}), it suffices to notice that for a black hole the scale of the Poincaré-breaking is the Planck scale, $f_P = M_P$. 
  Likewise, it is clear that the relation between the Bekenstein-Hawking entropy (\ref{BHent}) and the quantum coupling of graviton evaluated at the scale $R$, $\tilde{\alpha} = (R M_P)^{d-2}$, satisfies the relation, as it was originally pointed out in \cite{Dvali:2011aa}. The bound described by \eqref{coupling_bound} and \eqref{Areaf} is therefore a more fundamental bound, which encompasses in itself the area law, for which the Bekenstein-Hawking entropy represents a particular case. 
    
  We note that in many instances, the saturation of the bounds (\ref{coupling_bound}), (\ref{NBound}), \eqref{Areaf} also implies the saturation of the Bekenstein bound. However, the three former bounds appear more general, as they must be respected regardless of the latter.

    Now, whenever the bound \eqref{coupling_bound} 
   (and thereby (\ref{NBound}) and(\ref{Areaf})) is saturated, it turns out that all the properties that previously were thought to be exclusive to black holes also appear.  In particular, all saturons display the relations (\ref{eq:thermal_rate}). 
    
    This has been demonstrated by explorations of saturons in renormalizable theories in which calculations can be performed reliably
  \cite{Dvali:2020wqi, Dvali:2021jto, Dvali:2021ooc, Dvali:2021rlf, Dvali:2021tez}.
  In addition, it has been shown 
  \cite{Dvali:2021ofp, Dvali:2023qlk}  
  that saturons exhibit the relation between the maximal spin and the entropy 
  identical to black holes.

  The understanding of all aspects of the phenomenon of saturation is of fundamental importance because of the following reasons. 

 First, it allows us to understand the mysterious black hole properties in terms of a universal underlying phenomenon that goes far beyond gravity.
   Due to this universality, saturons in renormalizable theories can be used as test laboratories not only for explaining the known black hole
   properties but also for predicting new effects. 
   
   One example of a new phenomenon is provided by the so-called ``memory burden" effect 
   \cite{Dvali:2018xpy, Dvali:2020wft, Dvali:2018ytn}, which suggests that the information load carried by any system of enhanced capacity of information storage (such as a black hole), exerts a stabilizing back-reaction on it. 
   The stabilization of saturated solitons by the memory burden effect has been discussed in \cite{Dvali:2021rlf, Dvali:2021tez, Dvali:2024hsb}.
 Various potentially-observational implications of this effect have been discussed in the literature, but shall not be considered here \footnote{For early work on implications for primordial black holes see, \cite{Dvali:2020wft, Alexandre:2024nuo, Thoss:2024hsr, Dvali:2024hsb}.}.

   Next, the understanding of saturons in renormalizable theories allows us to shed light on the microscopic structure of a black hole. In particular, it provides a framework for testing the microscopic description of a black hole as a saturated coherent state of gravitons of occupation number $N = \tilde{\alpha}^{-1}$ 
  \cite{Dvali:2011aa, Dvali:2012en, Dvali:2013eja}.
 This picture, called the black hole's quantum $N$-portrait, explains the black hole properties microscopically. In particular, one can explicitly recover features of the Hawking radiation 
 (\ref{eq:thermal_rate}) from the quantum rescattering of the graviton constituents.
  The study of saturons in renormalizable theories 
  allows us to verify this correspondence,
  since saturons admit an explicit microscopic resolution in terms of coherent states of quanta.
  They can therefore demonstrate
  how the black hole-like features emerge from such states.  For example, the saturation of the entropy bound (\ref{NBound}) by a black hole is automatic within the $N$-portrait description  \cite{Dvali:2011aa, Dvali:2012en, Dvali:2013eja}. This feature 
  has been explicitly verified for other saturons 
  and shall also be the case in the present work.

   Finally, the study of saturons is important for the general approach of understanding the strong field non-perturbative semiclassical phenomena in terms of fully quantum multi-particle effects. This program was initiated in 
 \cite{Dvali:2011aa, Dvali:2012en, Dvali:2012rt, Dvali:2012wq, Dvali:2013eja, Dvali:2013vxa, Dvali:2017eba, Dvali:2014gua, Dvali:2024dzf}
for strong field gravitational systems such as black holes and de Sitter, and it was generalized to non-gravitational soliton and instanton processes in a series of papers \cite{Dvali:2019jjw,Dvali:2019ulr,Dvali:2020wqi,Dvali:2021ooc,Dvali:2021rlf,Dvali:2021tez,Dvali:2022vzz, Berezhiani:2020pbv, Berezhiani:2021gph, Berezhiani:2021zst, Berezhiani:2023uwt, Berezhiani:2024pub, Dvali:2017ruz}.     
 
    The main focus of the present paper will be on retrieving the Hawking-like evaporation rate \eqref{eq:thermal_rate} for a specific example of a saturon in a $1+1$-dimensional theory. 
     This decay rate has been reproduced previously 
in \cite{Dvali:2021ooc, Dvali:2021rlf, Dvali:2021tez} in $1+1$ and $3+1$ dimensional examples of saturons, using the picture of the 
$S$-matrix scattering of constituents. 
In \cite{Dvali:2021ooc}, the role of the saturon was 
played by a bound-state of the Gross-Neveu model \cite{Gross:1974jv}, whereas in 
\cite{Dvali:2021rlf, Dvali:2021tez} the saturon is represented by a stabilized version of the vacuum bubble of a spontaneously broken $SU(N)$-symmetry \cite{Dvali:2020wqi}. 

 The example considered in the present paper 
 is a $1+1$-dimensional version of the above-type vacuum bubble. 
 The novelty is that we shall perform 
 computations both in a semiclassical 
 as well as in a fully quantum regime, and 
 demonstrate the matching. 
  The semiclassical computation can be regarded  
  as a non-gravitational analog of the Hawking
  analysis. In this treatment, the saturon is taken as a background which sources quantum particles in the zero-back-reaction limit. In this regime, the particle creation is a vacuum process very similar to Hawking's semiclassical effect in a black hole. Thanks to the simplicity of the model, some results can be obtained analytically. 

 On the other hand, in the fully quantum description, 
 the decay rate is obtained as a result of 
 the re-scattering and decay of the constituent quanta. For the vacuum bubble, these constituents represent the Nambu-Goldstone modes of the spontaneously broken $SU(N)$-symmetry localized within the bubble interior. 
   On the gravitational side, this analysis 
    represents a counterpart of the black hole decay described within the quantum $N$-portrait 
    \cite{Dvali:2011aa}.

In the proper limit, the two approaches match, as expected. The presented saturon model thus can be viewed as a calculable laboratory that allows to explicitly interpolate between the Hawking-type semiclassical and quantum-corpuscular regimes. The advantage of the corpuscular picture is that, being more fundamental, it allows us to go beyond the leading-order approximation and estimate the corrections to the semiclassical behaviour. 

In particular, it allows us to explicitly track the 
process of the information retrieval and to 
establish the required time-scales, which are in accordance with the previous estimates 
\cite{Dvali:2020wqi, Dvali:2021ooc, Dvali:2021rlf, Dvali:2021tez}. Within the quantum picture, these time-scales are finite but go to infinity in the semiclassical limit. We thus explicitly observe, at the level of a calculable unitary theory, that seeming non-recovery of information is an artifact of the semiclassical approximation. 
\\  
  
	The paper is structured as follows: in the second section, we introduce the model and we discuss the bubble-like solutions which appear in its spectrum, with a particular focus on saturation. In the third section, we describe the setting of the semiclassical computation and we obtain the radiation spectrum of the bubble in the saturation limit. In the fourth section, we give the quantum description of the evaporation process. In the fifth 
    section, we interpret the results and estimate the corrections to the semiclassical behaviour. In the sixth section, we present the quantum derivation of the time-scale of the information retrieval. In the seventh section, we conclude.

	\section{A model of a saturon}
	\label{2}

\subsection{The classical soliton}
	We shall study the two-dimensional version of the model of a saturated vacuum bubble 
    originally introduced in \cite{Dvali:2020wqi}. 
    We shall focus on the stabilized version of the bubble derived in  \cite{Dvali:2021tez}. 
	The theory is defined by the following Lagrangian:
	\begin{align}
		\mathcal{L} &=\frac{1}{2} \operatorname{tr}\left[\left(\partial_{\mu} \Phi\right)\left(\partial^{\mu} \Phi\right)\right]-V[\Phi] \\
		V[\Phi] &=\frac{\alpha}{2} \operatorname{tr}\left[\left(f \Phi-\Phi^{2}+\frac{I}{N} \operatorname{tr}\left[\Phi^{2}\right]\right)^{2}\right] \, , \label{eq:matrix_potential}
	\end{align}
	where $\Phi$ is a scalar field in the adjoint representation of the $SU(N)$-group.
     Since we are in $1+1$ space-time dimensions, $\Phi$ is dimensionless. So is the 
     parameter $f$, whereas the parameter $\alpha$ has dimensionality of [mass]$^2$. 
      The dimensionless coupling, $\tilde{\alpha}$,  controlling the strength of the interaction,   is determined as 
     \begin{equation} \label{tilde}
     \tilde{\alpha} \equiv \frac{\alpha}{m^2} = \frac{1}{f^2} \,.
 \end{equation}
         
      In order to obtain saturated states, we shall work in the limit of large $N$ and large $f$ (weak coupling). Defining the 
      collective ('t Hooft) coupling as $N/f^2$, unitarity imposes the following upper bound on it, 
   \begin{equation} 
       N/f^2 
\lesssim
1 \,. 
    \end{equation}

    The potential \eqref{eq:matrix_potential} admits the $SU(N)$-symmetric vacuum as well as a number of degenerate vacua in which the $SU(N)$ symmetry is spontaneously broken down to various maximal little groups, 
    $SU(N-K) \times SU(K) \times U(1)$ with $0<K<N$. 
    For definiteness, we shall focus on two neighboring vacua: the $SU(N)$-symmetric vacuum in which the field vacuum expectation value (VEV) is zero and the vacuum in which the symmetry is spontaneously broken down to the $SU(N-1)\times U(1)$-subgroup:
    \begin{equation}
        \Phi_{\alpha}^{\beta} =\frac{f}{\sqrt{N(N-1)}} \operatorname{diag}((N-1),-1, \ldots,-1)\,.
    \end{equation}
    In the $SU(N)$-symmetric vacuum all excitations are gapped, with the 
    mass of the quanta given by  $m \sim \sqrt{\alpha} f$.  
    In the $SU(N-1)\times U(1)$-symmetric vacuum there exist $2N$ gapless Goldstone 
    species. 
       
  The theory admits vacuum bubble configurations interpolating 
 between the two vacua and separated by a bubble wall of thickness
 $\sim m^{-1}$. 
 
 Since we are in $1+1$-dimensions, a bubble represents a kink and an anti-kink
 separated by a certain distance $R$.  For infinite separation, the 
 kink (or anti-kink)  located at $x=0$  has the following form, 
\begin{equation} \label{kink}  
 \phi(x) = \frac{f}{2} \left ( 1 \pm \tanh(mx/2) \right)\,.
\end{equation} 
 The mass of a kink (or anti-kink) is given by, 
 \begin{equation} \label{kinkMass}
   E_{k} = \frac{1}{6}\frac{m^3}{\alpha}  = 
\frac{1}{6}\frac{m}{\tilde{\alpha}}\,.
 \end{equation}
  We consider the vacuum bubbles of the broken symmetry vacuum in 
 their interior, embedded in the symmetric one.  For $R \gg m^{-1}$ 
 the  asymptotic field values are $\phi(x) \simeq f$ for $|x| < R$ and $\phi(x) = 0$ for $|x| >  R$.  
 The interior of a bubble ($|x| < R$) houses $\sim 2N$ gapless Goldstone species.   
If no Goldstone modes are excited, the bubble at finite $R$ is unstable. 
The kink and anti-kink attract each other, and after some oscillations annihilate. 

  However, as shown in \cite{Dvali:2021tez}, the bubble can be stabilized by the Goldstone charge.  The theory admits stationary bubble solutions of the form: 

	\begin{gather}
	\label{eq:ansatz}	\Phi_{\alpha}^{\beta}=\left(U^{\dagger}V U\right)_{\alpha}^{\beta} \\
		U(t)=\exp \left[-i \theta^{a}(t) T^{a}\right]
	\end{gather}

	\begin{equation} \label{vev}
		V_{\alpha}^{\beta}(x)=\frac{\phi(x)}{\sqrt{N(N-1)}} \operatorname{diag}((N-1),-1, \ldots,-1)\,,
	\end{equation}

  where $T^a$ are the $SU(N)$ generators. 
  
    The bubble profile is easily investigated with the choice of $\theta^a$ as a uniform rotation in a single flavour, $\theta^a=\delta^a _1 \omega t$. Then, by plugging the ansatz \eqref{eq:ansatz} in the original Lagrangian, finding the bubble profile is reduced to the solution of the equation: 
    \begin{equation}
        \partial_x ^2 \phi +\mathcal{V}'(\phi)=0 \,,
    \end{equation}
     where the potential is given by
	\begin{equation} \label{potential}
		\mathcal{V}(\phi)=\frac{1}{2} \phi^{2}\left(\omega^{2}-\alpha(\phi-f)^{2}\right)\,.
	\end{equation}
 The stationary bubble profile corresponds to the solution that has $\phi(0) = \phi_0 \neq 0$ and asymptotes to $\phi(x) \rightarrow 0$ for $|x| \rightarrow \infty$.

 The proof of the existence can be obtained by following the reasoning given in \cite{Dvali:2021tez} for the $3+1$-dimensional version of the present model. The 
$1+1$-dimensional case considered here is simpler as certain characteristics can be obtained explicitly. 
 \begin{figure}[!ht]
     \centering
     \includegraphics[width=0.8\linewidth]{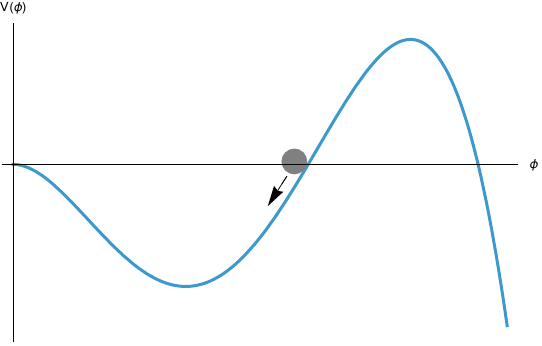}
     \caption{The effective potential for the $\phi$ field. The bubble solution corresponds to the field ``rolling" down from the zero value at ``time" $r=\infty$ to the finite value $\phi_0$ at ``time" $ r=0$, and back to $\phi=0$.}
     \label{fig:enter-label}
 \end{figure}

 If we think of $x$ as a time coordinate, the above equation is equivalent to the one satisfied by a classical particle with coordinate $\phi$ moving in the external potential (\ref{potential}). 
 Since in our parameterization $\phi$ is positive, we only consider the region $\phi >0$. The generalization of our analysis to a field defined on the entire real axis is trivial. 
 
 For $\dfrac{\omega^2}{\alpha f^2} \, < \, 1 $
 (or, equivalently, for $\omega^2 \, < \, m^2$),
the above potential has two maxima, $\phi = 0$ and $ \phi_{\text{max}} = \frac{f}{4} (3 + \sqrt{ 1 + \frac{8\omega^2}{\alpha f^2}} )$, separated by a minimum at $\phi_{\text{min}} = \frac{f}{4} (3 - \sqrt{ 1 + \frac{8\omega^2}{\alpha f^2}} )$. In addition to $\phi =0$, the potential  also becomes zero at the points $\phi_{0} = f \pm \frac{\omega}{\sqrt{\alpha}}$. 

In this regime, the existence of a stationary bubble solution is rather transparent. The ``particle'' starts at the ``time'' $x=-\infty$ with the initial position $\phi = 0$ and with zero initial velocity,
$\phi' =0$.  At some point it rolls down the hill, climbs on the other side of the slope and reaches the point $\phi_0  = f  - \frac{\omega}{\sqrt{\alpha}}$ with the zero velocity. 
It then bounces and repeats the excursion backward: rolls down, then up and reaches the initial maximum
$\phi = 0$ at $x= +\infty$. 

Thus, the bouncing point 
\begin{equation} \label{bounceP}
		\phi_0= f\left(1-\frac{\omega}{f\sqrt{\alpha}}\right)
	\end{equation}
determines the value of the field at the center of the bubble. 
The bubble size, $R$, can be defined as the characteristic ``time" interval required for 
the entire round trip. Each one-way journey represents a kink and an anti-kink, separated by the time interval that the field spends near the bouncing point $\phi_0$.  This time interval is controlled by the frequency $\omega$.

Notice that the time interval spent at the bouncing point  $\phi_0  = f - \frac{\omega}{\sqrt{\alpha}}$ goes to infinity for $\omega \rightarrow 0$, since in this limit the bounce-point coincides with the maximum:
$\phi_0 =  \phi_{\text{max}} = f$. Correspondingly, 
the particle spends an infinite time there before starting the return journey. That is, 
the time interval separating the onward and backwards transitions becomes infinite. 
In other words, in this limit, the bubble becomes a kink and an anti-kink, given by 
(\ref{kink}), at infinite separation,
$R \rightarrow \infty$.

    Thus, as in the $3+1$ case of \cite{Dvali:2021tez}, choosing a certain frequency $\omega$ for the Goldstone oscillations in the bubble interior fixes the size $R$ and the height $\phi_0$ of the bubble profile as in  
(\ref{bounceP}).     

    However, in contrast to the $3+1$-dimensional case \cite{Dvali:2021tez}, in the $1+1$ case for $\omega \rightarrow 0$ the size of the stationary bubble $R$ increases only logarithmically.  
 Namely, in the thin-wall approximation $m\gg\omega$ we have 
 \begin{equation}\label{bubble_size}
		R\sim \frac{1}{m}\ln\left(\frac{m}{\omega}\right) \,.
	\end{equation}
 The reason for the slow growth is that for $R \gg m^{-1}$, the attractive potential 
 between the kink and anti-kink is suppressed exponentially. 

	  Thus, in the thin-wall limit $\phi_0\simeq f$. Moreover, we have
    \begin{equation}
		R\omega\sim\frac{\omega}{m}\ln\left(\frac{m}{\omega}\right)\ll 1
	\end{equation}
	and 
	\begin{equation}
		\frac{\omega}{\phi_0}\ll \sqrt{\alpha} \,.
	\end{equation}
	For thick-wall bubbles, on the other hand,  we have $\omega\sim m$,  $R\omega \sim 1$ and $\phi_0 < f$.

   In the two-dimensional case, the bubble profile 
   $\phi_b(x)$ can be obtained analytically. 
 Due to the integral of motion, the bubble satisfies the first-order differential equation, 
\begin{equation} \label{eqB}
 d_x \phi_b = {\sqrt{-2\mathcal{V}(\phi_b)}}\,.
 \end{equation} 
which can be integrated explicitly to obtain
\footnote{In a different context, this solution was previously obtained by Juan Sebastián Valbuena Bermúdez. We thank him for sharing it.}: 
 \begin{equation} \label{Bprofile}
    \phi_b(r)= \frac{m-\omega}{\sqrt{\alpha}}\left(\frac{m+\omega}{m+\omega \cosh \left(\sqrt{m^2-\omega^2} r\right)}\right)\,.
 \end{equation}

 \begin{figure}[!ht]
     \centering
     \includegraphics[width=0.8\linewidth]{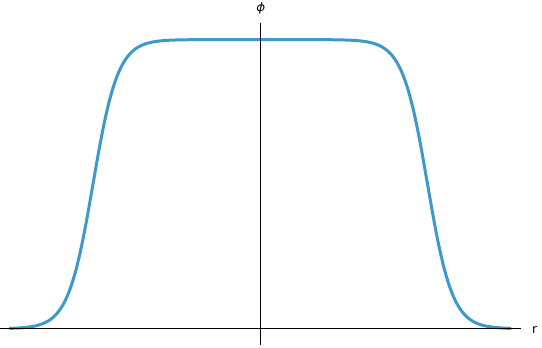}
     \caption{The bubble profile for the $\phi$ field.}
     \label{bubble_plot}
 \end{figure}

Correspondingly, the energy of the bubble is
\begin{equation} \label{Ebb}
 E_b = \int_{-\infty}^{\infty} (d_x \phi_b)^2 \,.
 \end{equation} 
  The analytic expression for the energy is readily available:

\begin{align}
 E_b = -\frac23mf^2\left(1-\frac{\omega^2}{m^2}\right)^{3/2}+mf^2\sqrt{1-\frac{\omega^2}{m^2}}+\nonumber\\+\frac{\omega^2f^2}{m}\ln \left(\frac{m}{\omega}-\frac{m}{\omega}\sqrt{1-\frac{\omega^2}{m^2}}\right)\,.
\end{align}

 In the thin-wall limit ($\omega/m \rightarrow 0$) this energy is equal to twice the energy of a static  kink (\ref{kinkMass}), 
\begin{equation} \label{EbbTW}
 E_b \simeq 2E_k \simeq \frac{1}{3} \sqrt{\alpha} f^3 =
 \frac{1}{3}  
 \frac{m}{\tilde\alpha}\,.
 \end{equation} 
However, order of magnitude wise it remains the same also in the thick-wall limit $\omega \sim m$. 

The $SU(N)$-charge of the bubble 
is given by
\begin{equation} \label{Qcharge}
 Q = \int dx (\phi_b(x))^2 \, \omega\,.
 \end{equation} 
 In the thin-wall approximation, the bubble size \eqref{bubble_size} can be 
 found by extremizing the $R$-dependent potential energy 
 \begin{equation}\label{thin-wall_potential}
 V(R) = \frac 12Rf^2\omega^2-2E_ke^{-mR},
 \end{equation} subject to the fixed charge constraint, which in thin-wall regime implies  $f^2R\omega \simeq Q$.  
 The second term in the potential comes from the attractive force between a 
 kink and an anti-kink.

\subsection{The Goldstone modes}

The bubble supports some gapless excitations, which can be interpreted 
as the Goldstone bosons of spontaneously broken 
space-time and internal symmetries.

In particular, the spectrum of the bubble excitations contains a gapless translation modulus, $\eta(t)$, which corresponds to infinitesimal translations of the bubble without a change in its shape. 
Correspondingly, the wavefunction profile of the translation modulus is given by the $x$-derivative of the bubble profile, 
\begin{equation} 
 \phi_b(x + \eta(t)) =  \phi_b(x) + 
 (\partial_x \phi_b(x)) \, \eta(t)\ \, + \, ...\,.
 \end{equation} 
 This modulus represents the Goldstone boson of a spontaneously broken translational symmetry.
 Notice that the norm of this degree of freedom, 
 \begin{equation}
 {\mathcal N}_{\eta} = \int_{-\infty}^{\infty} (\partial_x \phi_b)^2 \,,
 \end{equation} 
 coincides with the energy of the bubble. 
 At the same time, together with the bubble size, it determines the order 
 parameter of Poincaré-breaking in the following way,  
 \begin{equation} 
 f^2_P = {\mathcal N}_{\eta} R \, \sim \, 
 \frac{(mR)}{\tilde\alpha} \, = \, (mR) \, f^2 \,.
 \end{equation} 
Notice that for a thick-wall bubble, where $(mR) \sim 1$, the scales of 
spontaneous breakings of Poincaré and $SU(N)$-symmetries become equal. This shall have a deep physical meaning for determining the entropy saturation of the bubble and its similarity with the black hole entropy. 

 Let us now analyze the $SU(N)$ Goldstones. 
 In order to do so, let us parameterize, without loss of generality, the rotation of the bubble by one of the broken generators by the angle $\theta(x_{\mu})$. The effective action for this mode is: 
 \begin{equation} \label{Slinear}
  S_{\theta}= \int d^2x \phi_b^2(x) (\partial_{\mu} \theta)^2 \,.  
 \end{equation}
  Next, we perform the mode expansion: 
  \begin{equation}
  \theta(x,t) = \sum_k \Theta_k(x) \zeta_k(t)\,,
  \end{equation}
  where the modes satisfy 
  $\ddot{\zeta} = -k^2 \zeta$. The mode functions 
  satisfy (see also the discussion in Appendix \ref{BublGaus})
  
  \begin{equation} \label{GModeEQ}
  \Theta_k'' + 2\frac{\phi_b'}{\phi_b} \Theta_k' + 
  k^2 \Theta_k = 0 \,, 
\end{equation}  
where prime denotes $\partial_x$.

First, we notice that there exists a zero mode (gapless) with $k=0$ and a constant wavefunction,

\begin{equation} \label{Ngold}
\Theta_0 = \left (\int d x \phi_b^2(x) \right)^{-\frac{1}{2}} \equiv \frac{1}{\sqrt{{\mathcal N_{\theta}}}} \,.
\end{equation}

Notice that the norm of the Goldstone is connected to the bubble charge through the relation,  ${\mathcal N_{\theta}} = Q/\omega$.

 The above $k=0$-mode as well as the analogous ones for other broken generators are the Goldstone zero modes responsible for the degeneracy of the bubble in $SU(N)$ flavour space. 

Notice that there is also a growing solution 
with $k=0$  \cite{Dvali:2002fi}, 
\begin{equation}
 \tilde{\Theta}(x) = \int_0^x \frac{dy}{\phi_b^2(y)} \,,
\end{equation}
However, this mode is not normalizable, i.e.,  
$\int_0^{+\infty} dx \phi_b^2 \tilde{\Theta}_0^2 = \infty $.

 Let us remark that at finite $N$ none of the modes are exactly gapless. Rather, they have tiny gaps given by the scale 
\begin{equation} \label{gapzero} 
 \Delta \omega_0 \sim \frac{1}{NR} \,.
\end{equation} 
 This gap can be understood as the result of the spread of the wave-packet, due to the breaking of the symmetries of the Hamiltonian. 
  The first, universal feature, is the spread due to the breaking of Poincaré symmetry. 
 
  Indeed, in a quantum theory, any soliton with a localization radius 
 $R$ is represented by a wave-packet, which breaks 
 the translational symmetry of the Hamiltonian. 
  Due to this, over time it spreads into a superposition of solitons at different locations. For any saturated soliton, the spread-out time is given by 
 \begin{equation} \label{tspread} 
 \Delta t =  \frac{1}{\Delta \omega_0} \sim NR \,.
\end{equation} 
  This spread also translates into a spread of the mode functions of all the localized modes, endowing all the would-be zero modes with the minimal energy gap given by (\ref{gapzero}). 

  For the Goldstones of the internal symmetry, the origin of the gap comes from the spread in the internal space of the $SU(N)$-symmetry. Indeed, at finite $N$, the would-be degenerate ``vacua" obtained by the relative 
  rotations of the bubble solution are not truly orthogonal. 
  Correspondingly, the would-be Goldstones have gaps given by (\ref{gapzero}). 

   For saturated solitons, which are the thick-wall bubbles, the two spreads go hand in hand since $N$ controls both effects: the breaking of internal $SU(N)$ as well as of Poincaré symmetries. 
   Of course, for under-saturated solitons, the two gaps can be different and are controlled by the respective occupation numbers and localization radii. 

   In general, the presence of super-soft memory modes with gaps given by (\ref{gapzero}) is a universal feature of saturons, which can potentially be probed by the corresponding super-soft radiation of comparable frequency \cite{Dvali:2021jto}.  

    Notice that the existence of a small gap does not affect the features of saturation of the bounds  (\ref{coupling_bound}), 
  (\ref{NBound}) and (\ref{Areaf}), since 
 the entire diversity of microstates occupies the energy gap $\sim 1/R$. 
  As previously recognized in the context of the black hole $N$-portrait, the above features are fully shared by the microstates of a black hole (see e.g., \cite{Dvali:2012rt, Dvali:2015ywa}).

\subsection{Non-zero modes}

 Let us now mention some important non-zero modes. 
In addition to the zero mode, we can also identify a mode with a gap of order $\omega$ associated with the breathing of the bubble. It is the easiest to identify this mode for a thin-wall bubble. 
 As already discussed above, in this approximation we can think of the bubble as kink and anti-kink separated by a 
distance $R$. The configuration is stable since the attractive exponentially-suppressed force is balanced by the repulsion due to the $SU(N)$-charge. 
 The small oscillations of the kinks about the equilibrium point are described by the following 
 effective equation

\begin{equation} \label{EQdelta}
    E=2E_k\frac{\dot\delta(t)^2}{2}+\left. \frac12\frac{\partial^2 V}{\partial L^2}\right|_{L=R} \delta(t)^2\,,
\end{equation}
where the potential arises as the second-order expansion of the potential energy \eqref{thin-wall_potential}:
\begin{equation}
V(L) = \frac 12 Lf^2\omega^2  - 2 E_k  e^{-Lm},
\end{equation} around its minimum $\partial V/\partial L =0$, $L=R$ at fixed charge $Q$ which 
relates $\omega$ and $L$ as 
$\omega = Q/f^2 L$. 

 From here we get,  
\begin{equation}
2V''(R)=-m^2E_ke^{-Rm}=-mQ^2/f^2R^2=-mf^2\omega^2,
\end{equation}
and $V''(R)/E_k\sim \omega^2$.

  Therefore, the equation of motion for the canonically normalized $\delta$ at the 
  linear order is:
 \begin{equation}
 \ddot{\delta}(t)  +  \omega_{\delta}^2  \delta(t) = 0\,,
\end{equation}
with $\omega_{\delta} \sim \omega$.

\subsection{The quantum picture}

	As discussed in \cite{Dvali:2020wqi, Dvali:2021tez}, from a quantum perspective 
    the bubble represents a multi-particle bound state of coherent quanta held together by the collective effects. In a non-stationary bubble  
    discussed in \cite{Dvali:2020wqi}, no Goldstones are excited. In this case, the main constituents of the bubble coherent state are the quanta of the radial mode $\phi(x)$. 
    However, in the stationary bubble, such as the one constructed in \cite{Dvali:2021tez}
 and in the present paper, there are important constituents in form of the Nambu-Goldstone modes.    

  Let us discuss them separately.  The  main contribution to the bubble energy
  from the radial mode $\phi(x)$ comes from the walls,
  and is of order $E_\phi\sim \int \grad{\phi} ^2\sim  m\phi_0^2$. This 
  corresponds to the collective energy of quanta of wavelength $m^{-1}$ and 
  occupation number 
	\begin{equation}
		N_{\phi}\sim \frac{E_{\phi}}{m}\sim  \phi_0 ^2 \,.
	\end{equation}
	On the other hand, the energy associated with the Goldstone charge comes from Goldstone particles localized in the bubble interior, which oscillate coherently with the typical frequency $\omega$. Their contribution to the energy is of order $E_G \sim \int  \dot{\phi}^2 \sim \phi_0^2 \omega ^2 R$.  As in the case of the radial mode, 
    it represents the collective energy of quanta of frequency $\omega$ and 
    occupation number 
	\begin{equation}
		N_G \sim \frac{E_G}{\omega} \sim \omega R \phi_0 ^2 \,.
	\end{equation}
Notice that the relation 
(\ref{Qcharge}) tells us that the occupation number of Goldstones is equal to the charge of the bubble,
\begin{equation} \label{Qcharge1}
 N_G = Q \,.
 \end{equation} 
This is not surprising, since the charge is stored in excitations of the Goldstone field.

	From the previous analysis, we see that the ratio between the occupation numbers of Goldstones and particles of the radial  field $\phi(x)$ is 
	\begin{equation}
		\frac{N_G}{N_{\phi}}\sim \omega R \,.
	\end{equation}
	In the thin-wall limit, $R\omega \ll 1$. Consequently, the bubble is dominated by the particles of the radial field. On the other hand, for thick-wall bubbles $N_G\sim N_{\phi}$ and the particle constituents of the bubble are equally shared between 
    the radial field and the Goldstone modes. \\

\subsection{The microstate entropy of the bubble} 

     We shall now analyze the entropy content of the bubble,
    following  the counting given in \cite{Dvali:2020wqi, Dvali:2021tez} for 
    $3+1$-dimensional case. The counting for $1+1$-dimensional bubbles 
    is essentially unchanged. 

The microstate degeneracy of the bubble can be estimated in two equivalent ways. 

    On one hand, we can take the point of view of an observer who lives in the interior of the bubble. The vacuum of this observer is a Goldstone vacuum, with 
    $N_{sp} \simeq 2N$ gapless species of Nambu-Goldstone modes. Their total occupation number is $N_G$ which can be redistributed among different Goldstone species     
    without affecting the macroscopic structure of the bubble. This implies that the bubble can be described by an arbitrary pattern of $N_{sp}$ numbers, $|n_1, n_2, ...n_{N_{sp}} \rangle $ which sum to $N_G$. Counting the entropy then reduces to a combinatorics problem.\\

    The second point of view is the one of an exterior observer.  This observer lives in 
    a symmetric vacuum and sees the bubble 
   as a multiparticle state transforming in large tensorial representation 
   of the $SU(N)$ group.  Therefore, the entropy is set by the dimension of the representation.

 Both countings give the following microstate entropy for thick-wall bubbles  \cite{Dvali:2020wqi, Dvali:2021tez}: 
    \begin{equation}
        S\sim 2N \ln \left[\left(1+\frac{2N}{N_G}\right)^{\frac {N_G}{2N}}\left(1+\frac{N_G}{2N}\right)\right]\,.
     \end{equation}

   A quick analysis of the previous equation shows that, for the critical value of 't Hooft coupling $f^2=N$, the entropy of thick-wall bubbles saturates the bound \eqref{coupling_bound}: 
      \begin{equation} \label{SSS}
       S\sim 2N \sim f^2\,.
    \end{equation}
    Thus, thick-wall bubbles at the unitarity limit become saturons.
    The significance of this regime and its connections to unitarity are discussed more in depth in section \ref{sec:large_N}. 

 Since saturated bubbles are thick-wall bubbles, this will be the regime relevant for our analysis. However, we will also analyze the thin-wall regime, since it allows us
 to obtain precise analytical results. We shall then extrapolate our findings to 
  the thick-wall case.\\

 \subsection{Coupling to fermions}

  The stability of the stationary bubble can be understood as the result of it carrying the Goldstone charge. For the bubble to decay, the corresponding charge must be emitted to its exterior. 
  However, the exterior represents the $SU(N)$-invariant vacuum 
  in which all the charge-carriers are gapped. This creates an energy barrier, ensuring the stability of the bubble.  In this sense, a bubble represents a version of a non-topological soliton or a $Q$-ball \cite{Lee:1974ma,Friedberg:1976me,Coleman:1985ki,  Lee:1991ax, Cohen:1986ct, Kim:1992mm}. Various implications of these objects have been studied in the literature (see,e.g, \cite{Kusenko:1997ad, Dvali:1997qv,
Kusenko:1997si,Kusenko:1997vi, Volkov:2002aj}).  

  As discussed in \cite{Dvali:2021rlf, Dvali:2021tez, Dvali:2024hsb},  the stability of the bubble can be understood as a manifestation of the  ``Memory Burden" effect \cite{Dvali:2018xpy, Dvali:2024hsb}. 
   As already said, the essence of the phenomenon is that an information load carried by a system of high capacity of information storage stabilizes it against the decay. 
   The saturated bubble represents such a system, since the information encoded in the 
   Goldstone modes in the bubble interior is much cheaper 
   in energy as compared to the same pattern stored in the $SU(N)$-invariant vacuum.
 Correspondingly, the Goldstone pattern creates a memory burden that stabilizes the 
 bubble. 

 In order to enable the decay of the bubble, we shall couple it to a field that can extract the Goldstone charge from the bubble interior without paying the high energy price. We accomplish this by coupling the adjoint field $\phi$ to a massless fermionic field in the fundamental representation of $SU(N)$, $\psi_\alpha$, via the following terms in the original Lagrangian:

\begin{equation}
		\mathcal{L}=\mathcal{L}_\Phi +i\bar{\psi}^\alpha \slashed{\partial}\psi_\alpha -g \bar{\psi}^\alpha \Phi_{\alpha}^{\beta} \psi_\beta\,.
\end{equation}

 This offers an efficient decay channel for the bubble, and, as we shall see, it leads to the bubble evaporation. The dynamics of the particle creation process is in some aspects similar to the decay of a non-topological soliton with $U(1)$-charge discussed in \cite{Cohen:1986ct, Clark:2005zc}.  The  goal in the present case 
is the understanding of the universal features of the process in the regime of saturation
which, in particular, allows establishing a clear 
parallel with the decay of a black hole.

\subsection {The large N limit}\label{sec:large_N}
Before discussing the evaporation process, let us establish the unitarity constraints on the coupling constants at large $N$. \\
	
	First, let us identify the relevant dimensionless couplings. The dimensionless coupling, $\tilde{\alpha}$, controlling  a tree-level $2\to 2$ scattering of quanta of the field $\Phi$, is given by $\tilde{\alpha}$
    defined in (\ref{tilde}). 

      At the same time, the leading-order (planar) loop corrections are controlled by the 't Hooft coupling \cite{tHooft:1973alw} $\lambda := N/f^2$.  The perturbative unitarity constrains this coupling to satisfy $\lambda\lesssim1$.\\

     Analogously, the $2\to 2$ tree-level scattering among the fermionic particles mediated
     by the exchange of virtual $\Phi$'s is controlled by the 
     dimensionless coupling constant $g^2/{\alpha f^2}$.
 The corresponding 't Hooft coupling is defined 
 (and constrained) as $\beta := g^2N/\alpha f^2\lesssim1$.\\

	Now, in order to remain within the validity domain of the theory 
    while taking the limit $N\to \infty$, we must 
    respect the above unitarity constraints.
    The exact semiclassical limit of the theory is then described by the following double-scaling limit: 

    \begin{align}\label{eq:classical_limit}
        &f^2\sim N \to \infty \quad \frac{\alpha}{m^2}\sim \frac{g^2}{m^2}\sim \frac{1}{N} \to 0\\
        &m^2 \sim \alpha N \, = \, \text{finite} \,.
    \end{align}
    In particular, this implies that the quantity $g\phi_0$, which effectively controls the strength of the interaction between the bubble and the fermionic field, remains finite.
    
     Let us now consider the most interesting regime of saturation of the unitarity bounds. In this regime, the limit $N\to \infty$ and the requirement that we keep the bubble radius $R$ (and hence $m$) finite, completely determine the behaviour of the system.\\
     
	In particular, we have $\lambda\sim \beta\sim 1$, which in turn implies 
    \begin{equation}\label{eq:unitarity bound}
    f^2\sim N,\quad \alpha \sim \frac{m^2}{N},\quad g^2\sim \frac{m^2}{N}\,.
    \end{equation}
Taking into account (\ref{tilde}), 
it is clear that 
the entropy (\ref{SSS}) of the saturated bubble satisfies (\ref{coupling_bound}), (\ref{NBound}) 
as well as the area law (\ref{Areaf}). Of course, since we are in a $1+1$-dimensional theory, the area of the bubble is just a $c$-number of order one.

    \subsection{Analysis of the fermions-bubble system}
    
After the discussion on how the parameters of the theory are related to each other in the large-$N$ limit, we classify the different regimes of the interaction between the bubble and the fermions. The relevant scales that remain finite in the large $N$ limit are the bubble size $R$, its frequency $\omega$ and the effective mass that the fermions acquire inside the bubble $m_\psi\sim g\phi_0$. The latter at the unitarity bound behaves as  
\begin{equation}
    m_\psi \sim g\phi_0\sim m\left(1-\frac{\omega}{m}\right)=m-\omega \,.
\end{equation}
Also, for every value of the coupling, we have  $m_\psi\leq m$, with equality reached in the above limit for a thin-wall bubble.
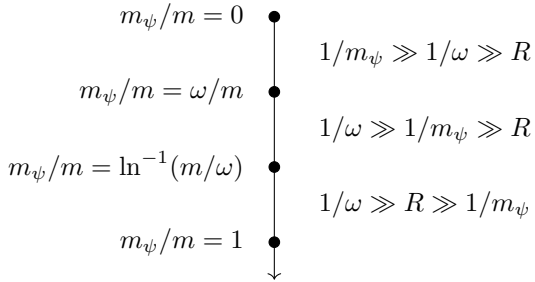
\begin{figure}[!ht]
    \centering
  
\begin{center}
\begin{tikzpicture}[scale=1]
\draw[<-] (0,0) -- (0,3.5);

\draw[fill=black] (0, 0.5) circle (2pt) node[left=8pt] {\( m_\psi/m=1 \)};
\draw[fill=black] (0,1.5) circle (2pt) node[left=8pt] {\(m_\psi/m= \ln^{-1}(m/\omega) \)};
\draw[fill=black] (0, 2.5) circle (2pt) node[left=8pt] {\(  m_\psi/m=\omega/m \)};
\draw[fill=black] (0, 3.5) circle (2pt) node[left=8pt] {\(  m_\psi/m=0 \)};

\node[] at (2,1) {$1/\omega \gg R\gg 1/m_\psi$};
\node[] at (2,2) {$1/\omega \gg 1/m_\psi \gg R$};
\node[] at (2,3) {$1/m_\psi \gg 1/\omega \gg R$};

\end{tikzpicture}
\end{center}

    \caption{Possible regimes for a thin-wall bubble.}
    \label{fig:thin-wall_regimes}
\end{figure}

\begin{figure}[!ht]
    \centering
  
\begin{center}
\begin{tikzpicture}[scale=1]
\draw[<-] (0,0) -- (0,2);

\draw[fill=black] (0, 0.5) circle (2pt) node[left=8pt] {\( m_\psi/m\lesssim 1 \)};
\draw[fill=black] (0, 2) circle (2pt) node[left=8pt] {\(  m_\psi/m=0 \)};

\node[] at (2,0.5) {$1/\omega \sim R\sim 1/m_\psi$};
\node[] at (2,1.3) {$1/m_\psi \gg1/\omega \sim  R$};

\end{tikzpicture}
\end{center}

    \caption{Possible regimes for a thick-wall bubble.}
    \label{fig:thick-wall_regimes}
\end{figure}
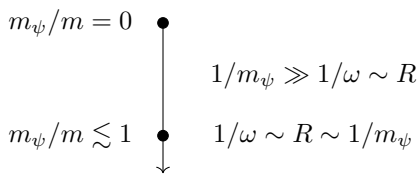
 Let us first consider bubbles in the parameter regime saturating the unitarity bound \eqref{eq:unitarity bound}:
	\begin{itemize}
		\item For thick-wall saturated bubbles, which lie in the regime $R\omega \sim 1$, the effective fermion mass can be smaller (but of the same order) than the energy of a Goldstone particle: $m_\psi \sim g\phi_0 \sim m-\omega \lesssim \omega$. This allows fermions to be on the verge of being created on-shell everywhere in the bubble.
		\item For thin-wall bubbles, $R\omega\ll 1$, the fermion mass inside the bubble is greater than the Goldstone frequency: $m_\psi \sim g\phi_0 \sim m \gg \omega$. Moreover, since $Rm_\psi \sim (m_\psi/m)\ln{(m/\omega)}\sim \ln{(m/\omega)}$, for a fermion of energy of order $ \omega$ the penetration 
        length inside the bubble, $\lambda_\psi \sim 1/m_\psi$, is much smaller than the bubble  size: $Rm_\psi\gg 1$.   In Section \ref{sec: quantum description}, we will see that this effect implies that fermion production can only happen close to the boundary of the bubble.
	\end{itemize}

	Now, in order to allow ourselves to remain safely within the regime below saturation, we take $g$ sufficiently small compared to its maximal value. Then the following regimes emerge:
    \begin{itemize}
        \item For thick-wall bubbles, by gradually decreasing $m_\psi$ we enter a regime in which $m_\psi \ll 1/R\sim \omega$. 
        \item For thin-wall bubbles, as $m_\psi$ decreases, two changes happen. First, when $m_\psi/m \sim \ln^{-1}(m/\omega)$, the fermion penetration length becomes larger than the bubble size: $Rm_\psi \sim 1$. Next,  by further decreasing $m_\psi$, the mass becomes also smaller than the Goldstone frequency $\omega$.
    \end{itemize}
This exhausts the possibilities that we can encounter regarding the relevant hierarchies of scales in the problem.
   
	\section{The semiclassical evaporation rate}
	
Having described our setup, we move on to obtain the semiclassical radiation spectrum of saturated bubbles.
	Namely, we aim to calculate the semiclassical particle creation for the $\psi$ field in the background of the bubble solution.  Of course, the bubble we are mainly interested in is a thick-wall saturated bubble. However, when the thick-wall conditions are satisfied, the profile function \eqref{Bprofile} of the bubble does not allow us to obtain an analytic solution for the fermionic mode functions. Therefore, we focus first on thin-wall bubbles where we can approximate the bubble profile by a rectangular function. In this regime, we can obtain the radiation spectrum analytically. We shall then extrapolate the result towards the thick-wall saturated regime.\\
    
    In order to facilitate the process, we choose the bubble with a macroscopically occupied Goldstone mode:  $\theta^a=\delta^a _1 \omega t$. Moreover, we shall work in the strict $N\to \infty$ limit. As discussed in the previous section, the quantity $g\phi_0$, which controls the interaction between the bubble and the fermions, remains finite as $N \to \infty$. Henceforth, all quantities of our interest 
    are regular in this regime. 
    In this limit, the matrix part of the vacuum expectation value of the field inside the bubble \eqref{vev} takes the simpler form
    \begin{equation} 
		V^{(N\to\infty)}{} _{\alpha}{}^{\beta}(x)=\phi_b(x)\operatorname{diag}(1,0, \ldots,0)\,.
	\end{equation}

    In order to understand the process of particle creation, we need to look at the modes of the fermionic field in the space-time dependent background of our bubble. As usual, whenever a field is quantized in a time-dependent background, the resulting quadratic Hamiltonian becomes time-dependent, and the number operators for asymptotic particles at early and late times are nontrivially related by a Bogoliubov transformation. The corresponding 
    Bogoliubov coefficients encode the rate of the particle creation by the bubble.\\

    To be more precise, let us first denote the two-momentum components by $k_0=\epsilon$, $k_1=k$. Since for our choice of $U(t)$ there are only two flavors for which both $V$ and $U$ are nontrivial, the particle production will 
    be controlled by these two flavors, to which we shall restrict our attention. Moreover, unless stated otherwise, we are going to use a unique index $i$ which runs over the tensor product of the spinor and the flavor spaces. This is convenient, since our calculation will mix
    the two types of indices.  With this convention, we label with $i=1,3$ the two spinor components of the first flavor and with $i=2,4$ the two spinor components of the second flavor. \\
    The generic mode expansion reads: 
    \begin{equation}
        \psi = \sum _n c_n \mathbf{f}^{ n} (x,t) \,,  
    \end{equation}
    where $n$ is an index that runs over the eigenvectors of the quadratic operator describing the equation of motion of the fermion in the bubble background and $\mathbf{f}^{n}$ are the corresponding eigenmodes.\\

    Obviously, since at spatial asymptotic infinity the fermions do not feel the presence of the bubble, their mode expansion reduces to the one of the plane waves of a free theory.
    Then, in each of the aforementioned regions the expansion becomes
	\begin{equation}
    \begin{split}
        \psi = \int_0^\infty d\epsilon\sum_{i} \frac{1}{\sqrt{2\omega_k}}c^i_{L/R} \mathbf{u}^{ \epsilon} _{i} e^{ i(k^i (\epsilon)  x-\epsilon t)}
		+\\+ \int_{-\infty}^0 d\epsilon\sum_{i} \frac{1}{\sqrt{2\omega_k}} c^{\dagger i} _{L/R} \mathbf{u}^{ \epsilon} _{i} e^{ i(k^i (\epsilon)  x-\epsilon t)} \,,
    \end{split}
	\end{equation}	
where $\mathbf{u}_i$ are the polarization spinors, which take into account also the choice of direction in flavour space.\\

The asymptotic behaviour of the mode functions is what allows us to define our vacuum states at infinite past and infinite future. Indeed, even though the time-dependent background generates a 
time-dependent Hamiltonian and hence a 
time-dependent ground state, the asymptotic regions far from the bubble give us a preferred way to define the ground states for $t\to \pm \infty$.

To this end, we split the mode expansion into purely incoming and purely outgoing waves, depending on the momentum. In particular, on the left-hand side of the bubble, the incoming waves with positive frequencies will have positive momenta, and vice versa for the negative-frequency waves. The opposite conditions hold for the right-hand side region of the bubble. 

The reason behind this splitting is that, for any given normalizable wavepacket, the purely incoming modes are the ones responsible for the creation (or annihilation) of the wavepacket at $t=-\infty$. On the other hand, the purely outgoing waves will control the states at $t=+\infty$.\\
This allows us to define the asymptotic vacuum at $t=\pm \infty$ as the state annihilated by ladder operators associated with a positive momentum from the $x\to -\infty$ side and with a negative momentum from the $x\to + \infty$ side, namely, a state with no incoming particles from both sides. Analogously, we define the asymptotic vacuum at $t = +\infty$ as the state with no outgoing particles. \\
Of course, if the bubble were absent, these two vacua would coincide and would correspond to the usual Fock space vacuum of the theory. The presence of the bubble mixes the different modes; in particular, a mode associated with a purely incoming wave at $t=-\infty$ evolves in a superposition of incoming and outgoing waves at $t =+ \infty$. Also, the bubble rotation generated by the $SU(2) $ matrix \begin{equation}\label{eq:fermion_redefinition}
	U(t)= e^{i\omega t T_1}=W\begin{pmatrix}
		e^{i \frac{\omega}{2}t} & 0 \\
		0 & e^{-i \frac{\omega}{2}t} 
	\end{pmatrix}W^\dagger \,,
\end{equation} 
where $W$ is a matrix describing a change of the basis that makes $U$ diagonal and is responsible for the flavor mixing,  mixes modes with positive frequencies $\epsilon$ with the ones with frequencies $\epsilon - \frac{\omega}{2}$ and vice versa. This gives rise to a Bogoliubov transformation among the incoming and the outgoing ladder operators of the form
\begin{equation}
    c_i{} ^{\mathrm{out}}= \sum_j \alpha_{ij} c_j{} ^{\mathrm{in}} +\beta_{ij} c^{\dagger} _j{} ^{\mathrm{in}}\,,
    \label{eq:Bogo_transformation}
\end{equation}
and leads to particle creation for frequencies $|\epsilon| <\frac{\omega}{2}$.
This fully specifies the relation among the asymptotic vacua at $t=\pm \infty$ and allows to extract the number density of created particles.\\

In order to carry on the explicit computation, it was useful to reduce the problem of solving for the modes to a time-independent problem by performing the following field redefinition for the fermions: $\psi' = U \psi$. \\
The equation of motion for the modes
\begin{equation}
	i\slashed{\partial}\psi+gU^\dagger V(x)  U \psi =0 \,,
\end{equation}
 then becomes: 
 \begin{equation}
     -\gamma_0 \omega T^1 \psi + i \slashed{\partial}\psi + gV(x) \psi =0\,.
     \label{eq:time_independent_EOM}
 \end{equation}

In general, for an arbitrary profile $V(x)$ an analytic solution cannot be found. However, the situation simplifies in the thin-wall approximation, where the bubble profile can be approximated by a double Heaviside Theta function. Of course, such an approximation implies that the validity of the calculation is restricted to modes with momentum much smaller than the wall-thickness: $k\ll m$. In this regime, we were able to obtain the full solution for the fermionic mode functions analytically, by imposing the continuity condition of the mode profile at the boundary of the bubble. \\
The resulting particle spectrum is obtained as a series expansion either in the regime $g\phi_0\ll \omega$ or in the opposite one, $g\phi_0 \gg \omega$. Some details of the calculations are included in the Appendix \ref{app: 1}. \\

  As discussed in Section \ref{2}, the regime relevant for saturation
 is $g\phi_0\lesssim \omega$. 
 Therefore, we can consider the evaporation process for a thin-wall bubble in this regime, and then extrapolate to a thick-wall, saturated bubble.
 Of course, for a saturated bubble, the thin-wall approximation is not precise, since we are in a thick-wall regime. However,
  for momenta $k < m$, order-of-magnitude wise, we can still trust the result in its parametric behaviour as a function of $g\phi_0$ and $\omega$.  
 
 In this regime, we obtain the density of created particles per frequency per unit volume in the leading order in a series expansion in the small parameter $g\phi_0/\omega$.  As expected, the radiation conserves the $SU(N)$ flavor and is democratic in particles and antiparticles. The radiation spectrum reads: \\
	\begin{equation}
		\rho_\epsilon =  \frac{g^2\phi_0^2 \sin^2(2R\epsilon)}{4\epsilon^2}\,,
	\end{equation}
	for $| \epsilon|<\frac{\omega}{2}$.\\
 
	 The fact that the first-order expression does not contain $\omega$ explicitly is an
    ``accident" of the first-order term of the series. 
    The second-order contribution, which was also obtained but will not be presented here, shows that  $\omega$ reappears in higher-order terms. \\
	
	In order to obtain the total particle density, we have to integrate over frequencies.
	
	\begin{align}
	& \Gamma= \int_0 ^{\frac{\omega}{2}} {\rm d}\epsilon ~ \rho_\epsilon  \nonumber \\
        &=\frac{g^2\phi_0 ^2 \left(2 R \omega  \text{SinIntegral}(2 R \omega )+\cos (2 R \omega )-1\right)}{4 \omega }\,.
        \label{eq:spectrum}
	\end{align}
	
 Extrapolating our results to the saturated regime, in which  $R\omega \sim 1$, all the terms in the parentheses become order one and we finally obtain,
	\begin{equation}\label{eq:saturon_rate}
		\Gamma \sim \frac{g^2\phi_0 ^2}{4 \omega}\sim \frac{1}{R}\,.
	\end{equation}

  We see that for a saturated bubble, the particle emission rate is given by the inverse size of the 
  bubble. This feature is strikingly similar to the 
Hawking-like evaporation rate of a black hole. 
  This similarity is fully in line with the 
  previously suggested connection between
  black holes and generic saturon states.
  In particular, the obtained decay rate is in agreement with the analytic results of saturon decays obtained in \cite{Dvali:2021rlf, Dvali:2021tez} for different models of saturons in the language of  QFT  $S$-matrix scattering amplitudes.

 In the opposite regime, relevant for a thin-wall bubble at the unitarity limit, the spectrum can be calculated in an expansion in the parameter $\omega/g\phi_0$. 
 In this regime, we have calculated the rate with the semiclassical method analogous to the one explained before. The result reads: 
\begin{equation}\label{eq:rate_thin_wall}
    \Gamma = \omega \tanh^2{\left(Rg\phi_0\right)}\,.
\end{equation}

Notice that this rate matches the corresponding rate of a thin-wall bubble at weak fermion coupling \eqref{eq:spectrum}, $\Gamma\sim g^2\phi_0^2R^2\omega$, in the parameter space where $\omega\lesssim g\phi_0\lesssim 1/R$. This signals a smooth transition between the two regimes.

Finally, even if a thin-wall bubble at the unitarity limit has a fermionic mass larger than the Goldstone frequency, $g\phi_0\gg\omega$, we can still try to extrapolate the evaporation rate to a thick-wall saturated bubble also from the equation \eqref{eq:rate_thin_wall}. This amounts to progressively tuning $g\phi_0\sim 1/R$ and at the same time $\omega\sim 1/R$. Up to order-one factors, we again recover the expected saturon evaporation rate \eqref{eq:saturon_rate}.

All considered, this discussion offers a clear picture of the bubble decay in its various regimes. In the next section, we shall see how the quantum description of the bubble fully reproduces all the features of its semiclassical behaviour in the proper limit.

\section{The quantum description of evaporation}\label{sec: quantum description}

 We shall now interpret the bubble evaporation process in a fully quantum language. 
 This language is similar to the one introduced 
 for the description of black hole decay in the context of $N$-portrait \cite{Dvali:2011aa} and 
 later applied  for describing the decay of saturon bubbles in \cite{Dvali:2021rlf, Dvali:2021tez}.
The general point is that the decay of the 
object is described as the depletion of the coherent 
state due to rescattering or decay of its constituent quanta. 
 
Applying this quantum description to the present picture, we describe the process as a rescattering of the bubble constituents into quanta of fermionic radiation. The dominant process is the decay of an excited Goldstone into a fermion-antifermion pair.

\begin{figure}[!ht]
\centering
\includegraphics[width=0.48\linewidth]{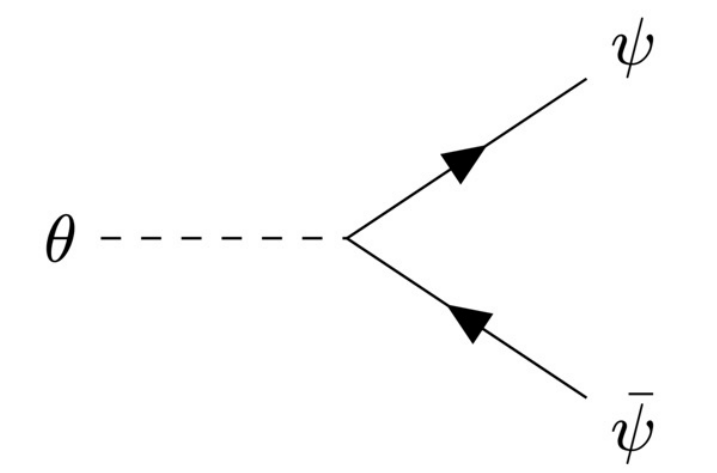}
\caption{The dominant channel that leads to the bubble evaporation.}
\label{fig:decay_channels}
\end{figure}

Of course, in an ordinary Goldstone vacuum, 
an on-shell massless Goldstone boson cannot 
decay. However, the situation in the case of a Goldstone localized within the bubble is fundamentally different. The Goldstone vacuum with a spontaneously broken $SU(N)$-symmetry 
exists exclusively in the interior of the bubble. 
In the asymptotic vacuum outside of the bubble, 
the $SU(N)$-symmetry is unbroken and no Goldstone 
modes are present. 

  Since the bubble breaks the space translation symmetry spontaneously, the dispersion relation of the localized Goldstone is very different from the 
  one of a free massless particle on the Poincaré invariant vacuum. 
Due to this, from the point of view of an asymptotic observer, a Goldstone mode
of frequency $\omega$ localized in the bubble represents an excited quantum level 
of the massive bubble.  
Such a state can get de-excited via a decay into a pair of free fermions without any kinematical obstruction. 

 Notice that in the double-scaling limit
(\ref{eq:classical_limit}), corresponding to the semiclassical regime, the back reaction on a bubble from any finite number of the acts of emission is zero.
That is, in this limit
the bubble effectively becomes an infinitely rigid time-dependent background which is eternal. 
  From the point of view of quantum fermion modes, this background represents a fully legitimate vacuum.  
Correspondingly, the quantum decay process of the constituent Goldstones can be described as a semiclassical process of particle creation in a time-dependent vacuum.
Hence, the particle creation rate 
 obtained in the semiclassical analysis can be 
 recovered as the large-$N$ limit (\ref{eq:classical_limit}) of the fully quantum picture of fermion production through the decay of the constituent Goldstone bosons.

Let us now proceed to the discussion of the quantum decay process. The decay rate of a Goldstone particle of the bubble is given by its characteristic frequency $\omega$ times the dimensionless quantum coupling $g/{\omega}$, which gives:
\begin{equation}
     \Gamma \sim \frac{g^2}{\omega}\,.
\end{equation}

This rate is enhanced by the Goldstone occupation number $N_G \sim \phi_0 ^2 R\omega$. Correspondingly, the total decay rate is given by: 
 \begin{equation}
     \Gamma^{tot} _g \sim \frac{g^2}{\omega} N_G\sim \frac{g^2}{\omega} (\omega R) \phi^2_0 \,.
 \end{equation}

However, this rate does not take into account the suppression due to the overlap of the wavefunctions between the fermionic field and the Goldstone field. Indeed, inside the bubble, where the Goldstone field is localized, the fermion acquires a mass $m_{\psi}\sim g\phi_0$.  For a low energy fermion, 
this restricts the penetration length of the  fermionic wavefunction  to  $\sim 1/ m_{\psi}$. 

 The physical effect of this restriction is that 
  a bubble Goldstone can efficiently decay 
  into fermions only in a region close to the boundary of the bubble, with a size of the order of the penetration length. 

This effect results in different suppression factors, depending on the regime in which we are operating. More details can be found in the Appendix \ref{app:wavefunction overlap suppression}. When the penetration length is much smaller than $1/\omega$ as well as the size of the bubble $R$, the small overlap induces a further suppression factor of the order $(\omega/R)(m_\psi)^{-2}$. As explained in section \ref{2}, this is the regime where $g\phi_0 \gg \omega$, relevant for thin-wall bubbles when the fermionic coupling is close to its unitarity bound. The resulting rate is 
\begin{equation} \label{QuantumR}
    \Gamma \sim \frac{g^2\phi_0^2}{\omega}\frac{\omega^2}{m_\psi^2}\sim \frac{g^2\phi_0^2 \omega}{m_\psi ^2}\sim \omega\,.
\end{equation}
We can compare this quantum rate with the one obtained via the semiclassical method \eqref{eq:rate_thin_wall} in the same regime. Indeed, taking into account that at the unitarity bound for thin-wall bubbles we have $g\phi_0 \sim m\gg1/R$, the semiclassical rate \eqref{eq:rate_thin_wall} matches the estimated rate (\ref{QuantumR}) obtained by the quantum analysis.

On the other hand, when the fermionic penetration length is of the same order or larger than the size of the bubble, $g\phi_0 \ll 1/R$, the dominant suppression in the overlap between the Goldstone and fermion wavefunctions comes 
from the localization width of the Goldstone 
mode which is given by the size of the bubble, $R$. 

In particular, in the thin-wall limit, where $R\ll 1/\omega$, the rate is further suppressed by a factor of order $R\omega$. This is relevant for the thin-wall bubbles at weak coupling, where $g\phi_0 \ll \omega$.
This regime is the one investigated by Eq \eqref{eq:spectrum}. Indeed, the decay rate matches with the one obtained by the semiclassical computation:
\begin{equation}
    \Gamma \sim \frac{(g\phi_0)^2}{\omega}( R \omega)^2 \sim \omega (R m_\psi)^2 \, .
\end{equation}

When evaluated at saturation, where $R\omega \sim 1$, this gives:
\begin{equation}
    \Gamma \sim \frac{g^2 \phi_0^2}{\omega}\sim \omega \sim \frac{1}{R}\,,
\end{equation}
 in accordance with the semiclassical computation.  \\

\subsection{An analytic regime for the de-excitation of the bubble Goldstones}

  Although the full analysis of all the regimes 
  of interests can be found in the Appendix \ref{app:wavefunction overlap suppression}, 
  we wish to present an analytic treatment 
  of a regime in which the de-excitation of the 
  bubble Goldstones can be made very transparent. 
   In this treatment, we assume that the bubble walls 
   are far apart, so that the system can 
   be treated as static for an exponentially long 
   time. The spectrum of the lowest lying modes can then be found exactly. In general, one needs to find the mode spectrum via the 
   analysis of the linear perturbations in the 
   bubble background. This will give a tower of 
   states characterized by the different frequencies
   $\omega_k$: 
\begin{align} 
  &\theta(x,t) = \sum_k \Theta_k(x) \zeta_k(t)=\sum_k \Theta_k(x) \, e^{i\omega_k t} \frac{\hat{a}_k}{\sqrt{\omega_k}} + h.c., \, \\ &\psi(x,t) =\frac{1}{\sqrt{L}} \sum_{k,\pm} \Omega_{k\pm}(x) \, \hat{b}_{k\pm} e^{i\omega_k t} \mathbf{u_{\pm}} + h.c.,  
  \end{align}
   Here,  $\Theta_k(x), \Omega_{k\pm}(x)$ are
   the mode functions for the Goldstone and the fermion field respectively, $\mathbf{u_{\pm}}$ are the fermion polarization spinors in the dimensionless normalization, and 
   $\hat{a}_k, \hat{b}_{k\pm}$ are the corresponding ladder operators. In the $\omega \to 0$ limit, the polarization spinors $\mathbf{u_{\pm}}$ correspond to 
   $\pm i$ eigenvalues of $\gamma_x$. Also, for 
   clarity, we are using a normalization for the modes within a box of finite size $L$, which later must be taken to infinity. 
   
    As we already discussed, for the  Goldstone mode with $\omega_k =0$ we have $\Theta_0(x) =1/\sqrt{\mathcal{N_\theta}}$,
   according to (\ref{Ngold}).

   \begin{figure}
       \centering
       \includegraphics[width=0.9\linewidth]{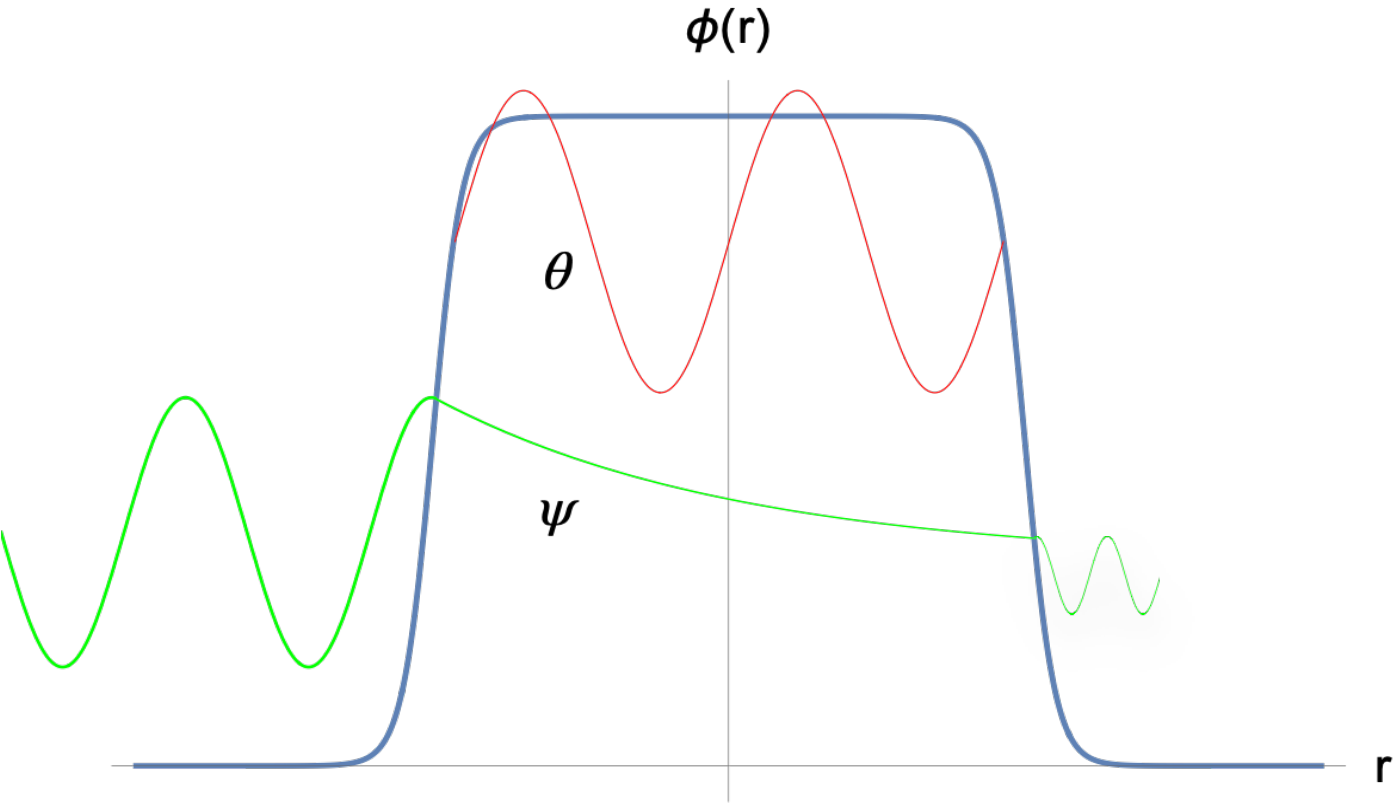}
       \caption{A graphical representation of the fermionic and Goldstone modes in the background of the bubble.}
       \label{fig:Bubble_modes}
   \end{figure}

 Once we know the profiles of the modes, we can 
    derive an effective interaction Hamiltonian 
    between the excited Goldstones and the fermions,  
    which allows for a simple computation of the transition rates.  

  In certain regimes, finding the mode functions can be significantly simplified. Namely, we shall be interested in the regime when the walls are far apart: 
   \begin{equation} \label{Osmall} 
   1/\omega \gg R , \, \, \omega \ll m_\psi \,,
   \end{equation}
and correspondingly focus on the Goldstone and 
fermion modes with frequencies $\omega_k \le  \omega$. 
  
 The profiles of such modes in the bubble vicinity are well-approximated by the profiles of 
 their $\omega_k=0$ counterparts. 
 In particular, in this limit,  the profiles of the fermion mode-functions $\Omega_{\pm} (x)$ 
 correspond to eigenfunctions of $\gamma_x$, $\gamma_x \psi = \pm i \psi$, and satisfy, 
 \begin{equation} \label{Fprofile} 
 \Omega_{\pm} (x) =  \Omega_{\pm} 
 (\mp \infty) \text{exp}\left ( \mp g \int_{\mp \infty}^x \phi_b(y) dy \right ) \,,
 \end{equation}
 where $\phi_b(x)$ is the bubble wall profile (\ref{Bprofile}). The mode for the Goldstone of frequency $\omega$ also tracks the profile for the zero mode, weighted by the bubble profile.
 
  Correspondingly, the overlap of the fermion and  the Goldstone 
 mode functions coming from their coupling in the 
 Lagrangian is (we drop the label $\omega$ because of the
 weak $\omega$-sensitivity): 
   \begin{align} \label{FBoverlap} 
   &\int dx \, g \, \phi_b(x) \, \Omega_{\pm}^2(x) \, = 
  \, 
  \int dx \frac{1}{2} d_x( \Omega_{\pm}(x)^2 ) \,\\
   &= \,   \Omega_{\pm}( \pm\infty)^2-\Omega_{\pm}( \mp\infty)^2\,.
   \end{align}
   The overlap is of order one for $1/m_\psi \ll R$, and is of order $m_\psi R$ for $1/ m_\psi \geq R$.
   
   We can now evaluate the decay process for the bubble perturbatively. Let us restrict to the modes of interest discussed above.
   
  Recall that we denote as
 \begin{equation}
 {\mathcal N}_{\theta} \,  = \, \int dx (\phi_b(x))^2 \sim 
 R \phi_0^2\,,
  \end{equation}  
the Goldstone normalization. Moreover, $\hat{b}_{1,2}$ and $\hat{a}_\omega$ are the fermion and Goldstone ladder operators, respectively.
The ladder operator $\hat{a}_\omega$ specifically refers to the Goldstone modes of the bubble with frequency $\omega$, which are the excited 
modes constituting the bubble. Similarly, 
 $\hat{b}_{1,2}$ refer to the number operators 
of the fermions of frequencies $\omega_1$ and 
$\omega_2$ into which the excited Goldstones decay.
 As we already said, due to 
 the condition (\ref{Osmall}), the profiles of such modes are very close to the ones 
 of the respective zero modes. 
 
 First, let us remark that the kinetic term for the Goldstone mode is normalized canonically:

   \begin{equation}
      H\supset \int dx \phi_b^2\Theta^2_0(x)\dot{\zeta}^2(t) = \omega \hat{a}^\dagger_\omega \hat{a}_\omega\,.
   \end{equation}
   
 We can now obtain the interaction vertex (we omit the spinorial structure for simplicity):
 \begin{eqnarray}
     \int dx g \phi_b \theta\bar{\psi}\psi= \int dx g \phi_b \Omega^2\frac{1}{L}b_1^+b_2^+a_\omega \frac{1}{\sqrt{\mathcal{N}_\theta \omega}}=\\\nonumber=\left(\Omega^2(\pm\infty)-\Omega^2(\mp\infty)\right)\frac{1}{2L}\frac{1}{\sqrt{R\omega\phi_0^2}}a_\omega b_1^+b_2^+\,.
 \end{eqnarray}

 In the regime $1/m_\psi \ll R$, the transition Hamiltonian among the properly normalized mode operators is:   
   \begin{equation} \label{FBoverlap1} 
\hat{H}_{\theta \rightarrow \psi}\, = \, \sum_{\omega_1, \omega_2 } \frac{1}{2 L \sqrt{\omega {\mathcal N}_{\theta}}} \, \hat{b}_{1}^{\dagger} \hat{b}_{2}^{\dagger} \hat{a}_\omega\,.
  \end{equation}
  In the regime $1/ m_\psi \geq R$, the operator above gets corrected by a factor of $m_\psi R$.

 The interaction Hamiltonian allows us to evaluate the matrix element from the initial state  $\ket{i} =\ket{N_G}$ to the final state
  $\ket{f} = \ket{N_G -1}\times \ket{1_b, 1_{b'}}$. Taking into account that the Goldstone occupation number is $N_G=R\omega\phi^2_0$, we obtain
     \begin{equation} \label{matrix} 
\bra{f} \hat{H}_{\theta \rightarrow \psi} \ket{i} \sim \sqrt{N_G}\frac{1}{L\sqrt{\omega R \phi_0^2}} \sim \frac 1L\,.
  \end{equation}
  Again, in the $1/ m_\psi \geq R$ regime, we get an additional $m_\psi R$ factor. 
  
  In order to obtain the rate, we need to integrate the squared matrix element over the phase space. Restoring the infinite space limit, $1/L \sim dp_i$, and integrating over the phase-space $\int dp_1\, dp_2 \delta(\omega-\sum_i\omega_i)$ gives us the total decay rate:
  \begin{equation}
      \Gamma\sim \begin{cases}
          \omega  & 1/m_\psi \ll R \\
          \omega R^2 m^2_\psi  &1/ m_\psi \geq R\,,
      \end{cases}
  \end{equation}
  in complete agreement with the previous sections.

\section{Understanding the spectrum}

  As we have seen, the bubble decay in the saturated regime captures the key feature of 
  black hole evaporation: both the decay rate 
  as well as the energies of emitted quanta 
  are fully determined by the inverse size of the bubble. 
   
  Yet, there are some visible differences from the spectrum of a black hole which we shall now clarify.   
 Namely, in the case of a black hole, the production of 
 particles with energy $ E\gg 1/R$ is exponentially suppressed but 
 not zero. Instead, in the case of a saturated stationary bubble, we are observing an exact suppression for particles of energy $ E >1/R$.

\begin{figure}[!ht]
    \centering
    \includegraphics[width=0.48 \linewidth]{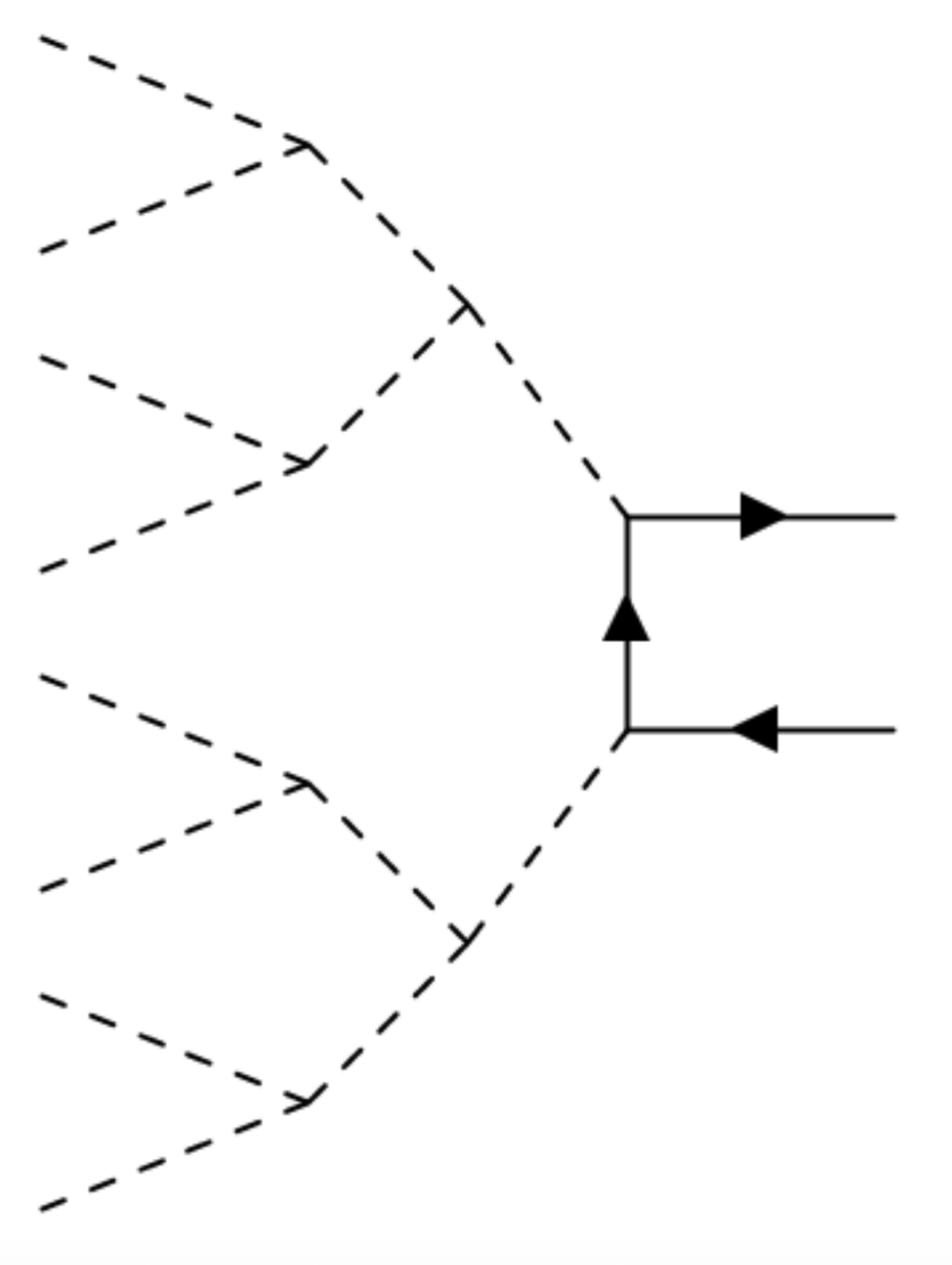}
    \caption{A would be $N \rightarrow 2 $ process depicting the annihilation of $N$ Goldstones into a pair of highly energetic fermions. Processes of this kind are forbidden by charge conservation.}
    \label{fig:N to 2}
\end{figure}
  
  There is a natural explanation for this.  
  In the case of a black hole,
 we can understand the suppression microscopically in the following way. 
 Thinking of a black hole as a coherent state of gravitons \cite{Dvali:2011aa}, we have 
 gravitons of all possible frequencies. The distribution is peaked 
 around the frequency $1/R$. However, much higher frequencies $E\gg 1/R$
 are also present albeit in exponentially suppressed numbers $\sim e^{-ER} $. 
 
 The production of energetic particles in the case of a black hole has at least two sources. The first is a direct decay of high-frequency constituent gravitons. Obviously, because of their suppressed number, the corresponding rate is exponentially suppressed. 
    
  The second source for production of asymptotic particles of energy $E \gg 1/R$ is the
     rescattering of a large number ($\Delta N \sim ER$)  of constituent gravitons
     into a quantum of energy $E$. Such rates are exponentially 
     suppressed as $e^{-ER}$ \cite{Dvali:2011aa, Dvali:2014ila, Dvali:2022vzz}. The suppression of the transitions with big changes in the 
     number of particles is a generic effect. For gravitons, it has been observed in the multi-graviton scattering amplitudes \cite{Dvali:2014ila, Addazi:2016ksu}. 
      The same feature is responsible for the suppressed production of solitons in particle scattering \cite{Brown:1992ay, Voloshin:1992rr, Argyres:1992np, Gorsky:1993ix, Libanov:1994ug, Libanov:1995gh, Son:1995wz, Dvali:2018xoc, Monin:2018cbi, Cornwall:1990hh, Goldberg:1990qk}.

     In the case of a saturated stationary bubble, none of these channels is available 
     in the semiclassical limit.  First, there are no constituent Goldstones 
     of frequencies different from $1/R$. Thus, the production of a high energy quantum 
     from the decay of a single constituent is impossible. 
      Secondly, each Goldstone carries a charge. Correspondingly, the annihilation 
      of $\Delta N = ER \gg 1$ Goldstones must produce a quantum of 
      a corresponding charge $\Delta N$. However, such quanta do not exist 
      in the theory. 
 Notice that,  since the bubble is not oscillating, the radial modes 
 carry no frequencies but only momenta. Due to this, the rescattering of radial 
 modes cannot increase the energy of the produced particle. 

  In this sense, the bubble shares some features of static topological solitons, which cannot decay due to exact same reason: no constituents of non-zero frequencies are available \cite{Dvali:2015jxa, Berezhiani:2024pub}. 

 The above reasoning, presented in the zero back-reaction limit (\ref{eq:classical_limit}),
 explains in simple terms the peculiarities of the bubble spectrum as well as some of its differences from the spectrum of a semiclassical black hole. Once we go away from this limit, the spectrum can get corrected due to various $1/N$-effects. For example,  since for finite-$N$ the bubble has a finite mass, the spectrum will be corrected due to a non-zero recoil. 
  
  Also, if we make the bubble oscillate radially around the classical equilibrium point, the radiation of modes with higher frequencies can appear, and will be exponentially suppressed \cite{Dvali:2022vzz}.

\section{time-scale of information retrieval}

  Let us now discuss the question of the time-scale of the information retrieval. Our reasoning and conclusions will be similar to the ones of the earlier papers  \cite{Dvali:2020wqi, Dvali:2021jto, Dvali:2021rlf, Dvali:2021tez, Dvali:2015aja} 
  addressing this issue.  
   First, within the validity of the semiclassical regime, it is impossible to retrieve any information from saturons, in full 
  analogy with the case of black holes. 
   In contrast, in the quantum treatment the time-scale of information retrieval is finite, but macroscopic.  
   
   This can be understood in very general terms of the necessary departure from a Hawking-like self-similar evaporation regime due to $1/S$-corrections to thermality \cite{Dvali:2015aja}.  
Such corrections are in full correspondence with the prediction of the black hole's $N$-portrait \cite{Dvali:2011aa}. However, the lesson from saturons allows us to understand the universality of this phenomenon, and to visualize it within a simple calculable framework.

For a saturated bubble, the information is stored in the $SU(N)$ quantum numbers of the Goldstones. 
 As classified in \cite{Dvali:2020wqi}, for an outside observer the information readout is possible in two ways, passive and active. 

 The passive regime implies the detection of 
 the outgoing radiation and the analysis of its 
 $SU(N)$-flavor content. The active regime implies the 
 readout of the bubble's flavor state by means of the
 scattering of external quanta at the bubble.

 Let us first discuss the passive method. 
 There are two factors that delay the 
 information read-out.  The first factor is the 
 weak coupling. For a saturated bubble, the emitted radiation has a frequency $\omega \sim 1/R$ and it interacts via the coupling constant $\tilde{\alpha}\sim 1/N \sim 1/S$.  The rate for resolving
 such a quantum in a detector is, 
 \begin{equation}\label{DETrate}
	 	\Gamma_{\rm det}  \sim  \frac{1}{R} \frac{1}{N^2}{N_{\rm det}} \,.  
	 \end{equation}  
 where, the factor $1/N^2$ comes from the coupling 
 and the $N_{\rm det}$ accounts for the capacity of the detector. At its maximum, this quantity is  
 $N_{\rm det} \sim N$.
 
 The rate (\ref{DETrate}) 
 fully resonates with the general black hole argument about the resolution of information encoded in $1/S$-corrections to thermality \cite{Dvali:2015aja}.
 It also matches the picture of black hole $N$-portrait according to which the information is encoded in radiation at the level of $1/N \sim 1/S$ effects \cite{Dvali:2011aa, Dvali:2012rt, Dvali:2012en, Dvali:2013eja}.

  However, the detection of a particle is just a beginning of the information read-out, since the identification of a single quantum gives only an exponentially small part of the bubble's information content. 
  
  Thus, even with a most sensitive detector, the start of the information retrieval is bounded from below by the time,
  \begin{equation}\label{Page}
	 	t \gtrsim \frac{1}{\Gamma_{\rm det}} \sim \, S \, R \,.  
	 \end{equation}  
 As already pointed out in \cite{Dvali:2020wqi,  Dvali:2021ooc, Dvali:2021rlf, Dvali:2021tez}, this time-scale is strikingly similar to the one proposed for a black hole by Page \cite{Page:1993wv}.

As discussed in \cite{Dvali:2021jto}, a universal way for understanding the information 
retrieval-time (\ref{Page}) is that it is equal to the time-scale (\ref{tspread}) given by the inverse gaps of the memory modes (\ref{gapzero}). It is clear that (\ref{tspread}), and correspondingly 
(\ref{Page}), is the minimal time required for processing the information encoded in any mode of frequency (\ref{gapzero}).

   The story with the active retrieval of information is different \cite{Dvali:2020wqi}. This way of information retrieval for black holes has not been discussed in the standard semiclassical context. In general, it is rather sensitive to the memory burden effect \cite{Dvali:2018xpy, Dvali:2020wft}.
 
 Following \cite{Dvali:2020wqi}, let us first discuss the active retrieval in a pre-memory-burden phase. This can be achieved by scattering the external radiation at a bubble. In our model, the role of the external probes can be played by the massive quanta of the adjoint field or by the massless fermion species that are coupled to it.  
 
  The scattering rate of a fermion of energy $\sim 1/R$ at a saturated bubble is of order one. This is because the suppression by the coupling $1/N$ is compensated by the occupation number of Goldstones $N_G\sim N$. However, this leading order effect does not reveal the bubble's $SU(N)$-flavor content. The resolution of the flavor quantum numbers is again suppressed by (\ref{DETrate}), amounting to the time-scale (\ref{Page}). 

  However, the above reasoning does not take into account the memory burden effect \cite{Dvali:2018xpy, Dvali:2020wft}. 
  Here, we shall limit ourselves by a qualitative discussion. The detailed analysis 
  of memory burden in saturated vacuum bubbles can be found in \cite{Dvali:2024hsb}.

  The memory burden of a black hole can be imitated by the bubble if we endow the fermions with bulk masses above the bubble's Goldstone frequency $\omega$.  This creates an energy difference between the bubble constituents and the asymptotic quanta of the same $SU(N)$-charge.
  
  Notice that within the composite framework of 
  black holes \cite{Dvali:2011aa, Dvali:2012en, Dvali:2017nis, Dvali:2018xpy} this gap is automatic, and goes up to $\sim M_P$. This is because in a black hole the information is encoded in high angular momentum modes, which outside of the black hole have very high frequencies but are gapless within the black hole.
  The prototype Hamiltonians capturing this effect have been constructed in \cite{Dvali:2017nis, Dvali:2018xpy}. 
  
  Once the bubble is in the 
  memory-burden phase, the emission is suppressed, and the active retrieval of information is the most effective way of reading it out. 
  From here on, the time-scale depends on the particularities of the system and can be different for black holes and other saturons (see \cite{Dvali:2024hsb}). 

 \section{conclusions}

 One of the key features of black hole physics is that the Hawking decay rate,  
 $\Gamma$, 
 as well as the characteristic energy of the emitted 
particles, $E$, are both set by the size, $R$, of the object,
 \begin{equation} \label{BHL}
    \Gamma \sim \frac{1}{R}\,, ~~~\, E \sim \frac{1}{R} \,.
\end{equation}
  At the same time, despite radiating the energy, the information stored in a black hole is not coming out until a macroscopically long time.   

On the other hand, in recent years it has been shown 
\cite{Dvali:2020wqi, Dvali:2021jto, Dvali:2021ooc, Dvali:2021rlf, Dvali:2021tez, Dvali:2021ofp, Dvali:2023qlk}  
that many features of a black hole are universally shared by the objects, the so-called saturons, that exhibit the maximal microstate degeneracy permitted by the QFT consistency.  The corresponding bound on 
the microstate degeneracy is given by three equivalent expressions (\ref{coupling_bound}), (\ref{NBound}) and (\ref{Areaf}). Within a consistent QFT, all bounds are saturated simultaneously. Thereby, they express different physical meanings of the same fundamental law. In particular, all three bounds are saturated by a black hole.  

It has therefore been suggested \cite{Dvali:2020wqi} that black hole features are not specific to gravity, but rather represent a universal outcome of the phenomenon of saturation.

In the present paper, we have demonstrated that the particular features (\ref{BHL}) are indeed exhibited by a class of saturated solitons in $1+1$ dimensions. 
The soliton of our interest represents 
a bubble of a vacuum with spontaneously broken 
$SU(N)$-symmetry embedded in the symmetric vacuum. 
  Due to the spontaneous breaking of the symmetry, 
 the bubble houses a large number ($\sim N$) of gapless Goldstone modes. These modes represent 
 the memory modes responsible for the maximal microstate degeneracy of the bubble. 
  Classically, the bubble is stabilized by the Goldstone charge. However, in the quantum theory, it can decay via the emission of bulk particles. 

  The black hole-like features of saturon decay 
  were previously demonstrated via the $S$-matrix analysis in \cite{Dvali:2021rlf}, where the saturon states have been shown to exist in Gross-Neveu theory \cite{Gross:1974jv}.  
 The case closer to the present paper is 
 the $S$-matrix analysis of the decay of the saturon bubbles in $3+1$-dimensions presented in \cite{Dvali:2021tez}.
 
 The lower-dimensional case studied in the present work allows us to derive the bubble decay more explicitly and to see a clear connection between 
 the semiclassical and quantum-composite pictures. 
 
 We analyzed the evaporation spectrum of a saturated bubble coupled to a massless fermionic field, using the semiclassical methods as well as the fully quantum picture of the bubble. As predicted from the general phenomenon of saturation, when the bubble enters the saturation regime, its decay exhibits the black hole features given by (\ref{BHL}).

The evaporation rate is obtained from the explicit computation of the Bogoliubov coefficients of the fermionic field at asymptotic times. 
We show that the semiclassical result matches the estimation from the fully quantum corpuscular description of the bubble. In this picture, the bubble evaporation is the result of the decay 
of the excited Goldstone modes, localized in the bubble interior, into the free bulk particles. 

The presented model allows to explicitly 
monitor the mechanisms of information storage and retrieval at the level of a calculable microscopic theory. As in the previous work on saturons
\cite{Dvali:2020wqi, Dvali:2021ooc, Dvali:2021rlf, Dvali:2021tez}, we observe the emergence 
of the time-scale 
of information retrieval (\ref{Page}). 
 We note that this time-scale reproduces 
 both the general estimate of 
 $1/S$-departures from thermality of black hole 
 evaporation \cite{Dvali:2015aja} as well as the 
 explicit predictions from the black hole $N$-portrait \cite{Dvali:2011aa} about the strength of such departures. 

It is striking (but perhaps hardly surprising) that the time-scale (\ref{Page}) fully matches the one previously introduced by Page \cite{Page:1993wv} in the context 
of a black hole. The saturon model gives a transparent microscopic meaning to this time-scale and shows why it extends far beyond gravity. 

At the same time, the saturon model also shows 
that the information retrieval process can be strongly influenced by the memory burden effect \cite{Dvali:2018xpy, Dvali:2020wft, Dvali:2018ytn}, which however we only discussed very briefly (for a recent detailed study of the effect in saturated bubbles, see \cite{Dvali:2024hsb}).

 The success of this quantum picture, which is rather transparent in non-gravitational theories, also provides an indirect support for the analogous quantum picture of a black hole as of a saturated coherent state of gravitons \cite{Dvali:2011aa}. 

The observed behaviour shows on an explicit example that saturons, even in a renormalizable $1+1$-dimensional theory that is drastically different from gravity, share the black hole properties such as the decay rate (\ref{BHL}). 
 This further reinforces the idea that these objects all belong to the same universality class of saturons. 

\section*{Acknowledgements}

We would like to thank Maximilian Bachmaier, Lasha Berezhiani and Juan Sebastián Valbuena Bermúdez for stimulating and fruitful discussions.

This work was supported in part by the Humboldt Foundation under the Humboldt Professorship Award, by the European Research Council Gravities Horizon Grant AO number: 850 173-6, by the Deutsche Forschungsgemeinschaft (DFG, German Research Foundation) under Germany’s Excellence Strategy - EXC-2111 - 390814868, Germany’s Excellence Strategy under Excellence Cluster Origins  EXC 2094 – 390783311 and 
the Australian Research Council under the Discovery Projects grants DP210101636 and DP220101721.     

\hskip 10pt

\noindent {\bf Disclaimer:} Funded by the European Union. Views and opinions expressed are however those of the authors only and do not necessarily reflect those of the European Union or European Research Council. Neither the European Union nor the granting authority can be held responsible for them.

\appendix

\section{derivation of the evaporation  spectrum }
\label{app: 1}
In this appendix, we provide some details of 
derivation of the radiation spectrum \eqref{eq:spectrum}. We begin by considering the Eq \eqref{eq:time_independent_EOM}. As explained above, it is sufficient to consider the two flavors that transform non-trivially under the generator of $SU(2)$-subgroup responsible for the bubble charge. 
Then, the evaporation dynamics is reduced to four relevant degrees of freedom.
In the thin-wall approximation, the Eq \eqref{eq:time_independent_EOM} splits into three free Dirac-like equations, two in the two asymptotic regions outside of the bubble and one in its interior. 

Notice that there is a time-dependent symmetry of the fermionic system in the background of the bubble. It is generated by a simultaneous time-translation and an $SU(2)$-rotation in the flavor space, $Q=\partial_t-\omega T_1$. Thus, we can label the modes according to the associated charge. This charge is exactly the energy $\epsilon$ of the modes, after the time-dependent fermion redefinition that we performed using \eqref{eq:fermion_redefinition}. On the other hand, translational invariance is broken by the bubble itself. Hence, from now on, it will be convenient to express the momentum $k$ of the modes, both asymptotically and inside the bubble, as a function of $\epsilon$, and not vice-versa, as it is customary. This will make the matching of the mode behaviour between the different regions more transparent.

In the two asymptotic regions, the modes describe a standard two-flavored massless fermion, where each flavor dispersion relation is modified by the field redefinition: 
\begin{equation}
    k_{1/2}(\epsilon)=\pm\epsilon \pm \frac{\omega}{2}\,.
    \label{eq:free_dispersion_relation}
\end{equation}

Inside the bubble, both fermions acquire an effective mass $m_\psi \sim g\phi_0$, and interact with the excited Goldstones. This leads to a flavor-dependent modified dispersion relation:

\begin{equation}
   k_{1/2}^2=\epsilon ^2-\left(\frac{m_\psi}{2}\right) ^2+\left(\frac{\omega}{2}\right) ^2\pm\frac{1}{2} \sqrt{4 \omega ^2 \epsilon ^2+m_\psi ^4-\omega ^2 m_\psi ^2}\,.
\end{equation}

The polarization vectors are modified accordingly. Note that for some frequencies the momenta of the modes become imaginary. This signals that the fermionic particles cannot propagate on-shell inside the bubble, as their energy would be smaller than their effective mass inside the bubble. Nevertheless, these off-shell modes must not 
be dismissed as they can connect the two asymptotic regions through tunneling.

The modes of Eq \eqref{eq:time_independent_EOM} hence are of the following form:

\begin{equation}
	\psi_{\epsilon}(x,t)=\begin{cases}
		\sum_{j} \tilde{c}^{j}	\tilde{\mathbf{u}}^{\epsilon} _{j} e^{\text{i}\left(\tilde{k}_{j}(\epsilon) x- \epsilon t\right)} & |x|<R \\
		\sum_{j} c_{L/R}^{j}	\mathbf{u}^{\epsilon} _{j} e^{\text{i}\left(k_{j}(\epsilon) x- \epsilon t\right)} & |x|>R \,,
	\end{cases}
\end{equation}
where quantities with a tilde refer to the solution inside the bubble, and the summation goes over both spin and flavor degrees of freedom. This expression holds for all frequencies $\epsilon$, both positive and negative. Note that the coefficients $c^i$ have to be interpreted either as creation or annihilation operators, depending on their frequencies. This is crucial for understanding how particle creation from the vacuum can occur. 

In order to obtain the full solution for the modes, we need to match the coefficients of the plane-wave expansion in the different regions for a given frequency. This allows us to obtain $c_{L}^{i}$ as a function of $c_{R}^{i}$
\begin{equation}\label{eq:smatrix}
	c^{L i}_{\epsilon} = S_{\epsilon} {} ^i{}  _ j c^{R\  j}_{\epsilon} \,.
\end{equation}
The complete expression for $S_{ij}$ is quite lengthy and we do not report it here. However, it contains all the necessary ingredients for extracting the coefficients of the Bogoliubov transformation \eqref{eq:Bogo_transformation}. Below, for illustrative purposes, we write explicitly the Bogoliubov coefficients for one of the outgoing ladder operators in the expansion for small $g \phi_0/\omega$.

As explained in the main body of the paper, the asymptotic vacuum can be defined at $t=\pm \infty$ by considering modes that propagate incoming or outgoing waves respectively. Once the full spectrum of modes is obtained by \eqref{eq:smatrix}, one can write the coefficients $c_i$ corresponding to outgoing waves as functions of the incoming ones as in \eqref{eq:Bogo_transformation}.

 A particle creation can occur when the Eq \eqref{eq:Bogo_transformation} mixes modes with positive and negative frequencies. If we undo our time-dependent fermionic field redefinition, it is clear from the Eq \eqref{eq:free_dispersion_relation} that this happens for $\epsilon\leq \frac{\omega}{2}$. Indeed, given that $\epsilon$ is conserved, the only mixing that can take place is among the modes that in the time-dependent basis have frequencies $\epsilon'$ and $\epsilon'-\frac{\omega}{2}$. Consequently, positive and negative frequencies overlap only in the range discussed above. Outside of this range, particle creation from the vacuum is excluded in the semiclassical limit. However, we still have a non-trivial scattering against the bubble. We report one of the Bogoliubov equations for the outgoing ladder operator associated with the flavor of the bubble (the others are similar and can be obtained as described above):

\begin{widetext}

\begin{equation}
\begin{split}
c_{1, \text{out}~\epsilon}^{L \dagger} =\; & 
c_{2, \text{in}~\epsilon}^{L \dagger} \left( \frac{i g \phi_0 \sin (R (2 \epsilon - \omega))}{2 \epsilon - \omega} \right) 
+ c_{3, \text{in}~\epsilon}^{R} \left( \frac{i g \phi_0 \sin (2 R \epsilon)}{2 \epsilon} \right) \\
& + c_{1, \text{in}~\epsilon}^{R \dagger} \left( 
    -\frac{2 i \epsilon^2 (g \phi_0)^2 e^{-i R \omega} \sin (4 R \epsilon)}{(\omega^2 - 4 \epsilon^2)^2} 
    + \frac{8 i \epsilon^3 (g \phi_0)^2 \sin (R \omega)}{\omega (\omega^2 - 4 \epsilon^2)^2} \right. \\
& \qquad 
    - \frac{\omega \epsilon (g \phi_0)^2 e^{-i R \omega} \cos (4 R \epsilon)}{(\omega^2 - 4 \epsilon^2)^2} 
    + \frac{\omega \epsilon (g \phi_0)^2 (\cos (R \omega) - 3 i \sin (R \omega))}{(\omega^2 - 4 \epsilon^2)^2} \\
& \qquad 
    + \frac{\omega^2 (g \phi_0)^2 \sin (4 R \epsilon) (\sin (R \omega) + i \cos (R \omega))}{2 (\omega^2 - 4 \epsilon^2)^2} 
    - \left. \frac{\omega^3 (g \phi_0)^2 e^{-i R \omega} \sin^2 (2 R \epsilon)}{2 \epsilon (\omega^2 - 4 \epsilon^2)^2} 
\right) \\
& + c_{4, \text{in}~\epsilon}^{L} \left( 
    \frac{2 \epsilon^2 (g \phi_0)^2 (\sin (R \omega) + i \cos (R \omega)) \sin (R (4 \epsilon - \omega))}{(\omega^2 - 4 \epsilon^2)^2} 
    - \frac{8 i R \epsilon^3 (g \phi_0)^2}{(\omega^2 - 4 \epsilon^2)^2} \right. \\
& \qquad 
    + \frac{\omega \epsilon (g \phi_0)^2 e^{-2 i R \omega} \left( e^{4 i R \epsilon} - e^{2 i R \omega} \right)}{(\omega^2 - 4 \epsilon^2)^2} 
    - \frac{2 i R \omega \epsilon^4 (g \phi_0)^2}{(\omega^2 - 4 \epsilon^2)^2} \\
& \qquad 
    + \frac{\omega^2 (g \phi_0)^2 e^{-2 i R (\omega + 2 \epsilon)} \left( 2 e^{2 i R \omega} - 3 e^{2 i R (\omega + 2 \epsilon)} + e^{8 i R \epsilon} \right)}{4 (\omega^2 - 4 \epsilon^2)^2} \\
& \qquad 
    + \frac{3 i R \omega^2 \epsilon^3 (g \phi_0)^2}{(\omega^2 - 4 \epsilon^2)^2} 
    + \frac{i R \omega^3 \epsilon^2 (g \phi_0)^2}{2 (\omega^2 - 4 \epsilon^2)^2} \\
& \qquad 
    + \frac{\omega^4 (g \phi_0)^2 (1 - e^{-4 i R \epsilon})}{16 (\omega^2 - 4 \epsilon^2)^2} 
    - \left. \frac{i R \omega^4 \epsilon (g \phi_0)^2}{4 (\omega^2 - 4 \epsilon^2)^2} + 1 \right)\,.
\end{split}
\end{equation}

\end{widetext}

 The precise computation of the particle spectrum follows directly from the expression for the modes contained in Eq. \eqref{eq:smatrix}. This has been obtained exactly in the thin-wall limit, and then expanded in series both for small $g \phi_0/\omega$ and in the opposite regime, in order to give the radiation spectra \eqref{eq:spectrum} and \eqref{eq:rate_thin_wall}.

\section{calculation of the wavefunction overlap suppression}\label{app:wavefunction overlap suppression}

In order to justify the suppression factors given in \ref{sec: quantum description}, we estimate the decay rate of a Goldstone particle $\theta$ of energy $\omega$ into a pair of asymptotic fermions of energy $\omega_1$ and $\omega_2$ and corresponding momenta $|p_1|=\omega_1$ and $|p_2|=\omega_2$. 
At tree level, the total decay rate reads: 
\begin{equation}
    \Gamma = \frac{1}{T}\int \frac{dp_1}{2\omega_1} \frac{dp_2}{2\omega_2}\frac{|\bra\theta g \int d^2 x \theta \bar{\psi}\psi\ket{\bar{\psi}\psi}|^2}{\bra{\theta}\ket{\theta}}\,.
\end{equation}

The Goldstone field is localized in the bubble of size $R$. 
Thus, the modes of the Goldstone particles populating the bubble are normalized by a factor $1/\sqrt{\omega R}$.
Let us denote by $\theta_\omega$ the mode function that corresponds to a Goldstone particle with frequency $\omega$.

On the other hand, the fermionic field in the background of the bubble is expanded in a continuum of modes. Again, let us denote by $\psi_i$ the mode that is associated with an asymptotic fermion of energy $\omega_i$. Note that if we were in an empty space, the process would not be kinematically allowed by momentum conservation. We will show below that this constraint gets relaxed in the presence of the bubble, and a larger kinematical window is allowed. 

Since the fermion is massive inside the bubble but massless outside, for energies smaller than the fermionic mass the modes will resemble a plane wave outside the bubble, and will be exponentially decaying inside with the penetration length $l_p\sim 1/m_\psi$. Indeed, inside the bubble the modes behave like $\psi_i(x) \sim \exp\left( ix\sqrt{\omega^2-m_\psi ^2}\right)$. 
Notice that, since the modes are oscillatory outside of the bubble, they are normalized with the usual continuum normalization. Namely, for each of them we have $<\psi_i|\psi_i> =(2\pi)\omega_i\delta(0) =(2\pi)\omega_iV$, where $V$ is the volume of the full space, and it should be understood as $\frac1V\sim dp$. 
When the energy $\omega$ exceeds the fermionic mass induced by the bubble, the fermion starts propagating on-shell also inside the bubble, with an effective momentum of order $k\sim \sqrt{\omega^2-m_\psi^2}$. 

Given the premises, we can estimate the decay rate of a Goldstone into a fermion pair. Summing over polarizations, we obtain: 
\begin{equation}
    \Gamma \sim g^2 \frac{1}{R\omega}  \int\frac{dp}{\omega_1}\frac{dq}{\omega_2}  \left|\int dx \theta^*_{\omega}\bar{\psi}_{1}\psi_{2}\right|^2 \omega_1\omega_2 \delta(\omega-\sum_i \omega_{i})\,.
\end{equation}
 The wavefunction overlap factor can be estimated depending on the regimes. In the thin-wall limit, $R\omega\ll 1$,  when $m_\psi\gg\omega$, the integral is approximately independent of the value of the asymptotic momenta \begin{align}
     \left|\int dx \theta^*_{\omega}\bar{\psi}_{1}\psi_{2}\right|^2\sim & \left|\int_{0}^{R} dx e^{i \omega x}e^{-2 m_{\psi}x}\right|^2 \\
     \sim & \begin{cases}
         \frac{1}{m^2_\psi} &  m_{\psi} \gg 1/R \\
         R^2 & m_\psi \ll 1/R\,.
     \end{cases}
 \end{align}
 The $\omega_i$-dependence of the wavefunction-overlap factor arises in the form of the $\omega/m_\psi$-corrections, which are contributing to the phase space integral. Energy conservation ensures that they remain suppressed compared to the leading-order contribution.
 When $\omega\geq m_\psi$, the fermions can be produced on-shell. The wavefunction overlap again is: 
 \begin{equation}
     \left|\int dx \theta^*_{\omega}\bar{\psi}_{1}\psi_{2}\right|^2\sim \begin{cases}
            R^2 & \omega \ll 1/R \\
            \frac{1}{\omega^2}\sim R^2 &  \omega\sim 1/R \,.
     \end{cases}
 \end{equation}
 Thus, the rate for a single particle scattering is of the order: 
 \begin{equation}
     \Gamma \sim \begin{cases}
         \frac{g^2}{\omega}\frac{1}{m_\psi^2}\frac{\omega}{R} & m_\psi \gg 1/R  \\
       \frac{g^2}{\omega}(\omega R) & m_\psi \ll 1/R \,. 
     \end{cases}
 \end{equation}
 Taking into account the occupation number of Goldstone particles, this leads to the results of \ref{sec: quantum description}.
 \section{Modes of a bubble with Gaussian profile}
 \label{BublGaus}
It can be instructive to consider modes of a bubble with the Gaussian profile $\phi_b^2(x) = e^{-(x\omega)^2/2}$. Although such a bubble is not a solution of the presented model, it can be viewed as an auxiliary construct for gaining 
an intuition about the Goldstone mode structure, which is exactly solvable.
The equation for the mode functions (\ref{GModeEQ}) takes the form \cite{Dvali:2002fi}, 
\begin{equation}
  \Theta_k'' + x\omega^2 \, \Theta_k' + 
  k^2 \Theta_k = 0 \,.  
\end{equation}  
which is solved by Hermite polynomials $H_n(\omega x/\sqrt{2})$ with $k = \omega \sqrt{n}$. 
Also, notice that in this case all the modes would be localized. Of course, in the present case the bubble profile is not a Gaussian, but this gives a qualitative idea about the structure of lowest lying modes. Notice that there exists an upper bound $k^2 <  m^2 -\omega^2$ on frequencies of the localized modes. This can be obtained by examining the equation for the bubble profile for $x \rightarrow \infty$, which takes the form: 
\begin{equation}
  \Theta_k'' - 2\sqrt{m^2-\omega^2} \, \Theta_k' + 
  k^2 \Theta_k = 0 \,.  
\end{equation}  
 For $k^2 <  m^2 -\omega^2$, this is solved by 
  $\Theta_k \rightarrow { e}^{Ax}$ with $A = \sqrt{m^2-\omega^2} (1 - \sqrt{1-k^2/(m^2-\omega^2)})$. The normalizable localized modes exist only with $k^2 <  m^2 -\omega^2$.

\bibliography{references}

\begin{thebibliography}{67}%
\makeatletter
\providecommand \@ifxundefined [1]{%
 \@ifx{#1\undefined}
}%
\providecommand \@ifnum [1]{%
 \ifnum #1\expandafter \@firstoftwo
 \else \expandafter \@secondoftwo
 \fi
}%
\providecommand \@ifx [1]{%
 \ifx #1\expandafter \@firstoftwo
 \else \expandafter \@secondoftwo
 \fi
}%
\providecommand \natexlab [1]{#1}%
\providecommand \enquote  [1]{``#1''}%
\providecommand \bibnamefont  [1]{#1}%
\providecommand \bibfnamefont [1]{#1}%
\providecommand \citenamefont [1]{#1}%
\providecommand \href@noop [0]{\@secondoftwo}%
\providecommand \href [0]{\begingroup \@sanitize@url \@href}%
\providecommand \@href[1]{\@@startlink{#1}\@@href}%
\providecommand \@@href[1]{\endgroup#1\@@endlink}%
\providecommand \@sanitize@url [0]{\catcode `\\12\catcode `\$12\catcode
  `\&12\catcode `\#12\catcode `\^12\catcode `\_12\catcode `\%12\relax}%
\providecommand \@@startlink[1]{}%
\providecommand \@@endlink[0]{}%
\providecommand \url  [0]{\begingroup\@sanitize@url \@url }%
\providecommand \@url [1]{\endgroup\@href {#1}{\urlprefix }}%
\providecommand \urlprefix  [0]{URL }%
\providecommand \Eprint [0]{\href }%
\providecommand \doibase [0]{http://dx.doi.org/}%
\providecommand \selectlanguage [0]{\@gobble}%
\providecommand \bibinfo  [0]{\@secondoftwo}%
\providecommand \bibfield  [0]{\@secondoftwo}%
\providecommand \translation [1]{[#1]}%
\providecommand \BibitemOpen [0]{}%
\providecommand \bibitemStop [0]{}%
\providecommand \bibitemNoStop [0]{.\EOS\space}%
\providecommand \EOS [0]{\spacefactor3000\relax}%
\providecommand \BibitemShut  [1]{\csname bibitem#1\endcsname}%
\let\auto@bib@innerbib\@empty
\bibitem [{\citenamefont {Bekenstein}(1973)}]{Bekenstein:1973ur}%
  \BibitemOpen
  \bibfield  {author} {\bibinfo {author} {\bibfnamefont {Jacob~D.}\
  \bibnamefont {Bekenstein}},\ }\bibfield  {title} {\enquote {\bibinfo {title}
  {{Black holes and entropy}},}\ }\href {\doibase 10.1103/PhysRevD.7.2333}
  {\bibfield  {journal} {\bibinfo  {journal} {Phys. Rev. D}\ }\textbf {\bibinfo
  {volume} {7}},\ \bibinfo {pages} {2333--2346} (\bibinfo {year}
  {1973})}\BibitemShut {NoStop}%
\bibitem [{\citenamefont {Bekenstein}(1981)}]{Bekenstein:1980jp}%
  \BibitemOpen
  \bibfield  {author} {\bibinfo {author} {\bibfnamefont {Jacob~D.}\
  \bibnamefont {Bekenstein}},\ }\bibfield  {title} {\enquote {\bibinfo {title}
  {{A Universal Upper Bound on the Entropy to Energy Ratio for Bounded
  Systems}},}\ }\href {\doibase 10.1103/PhysRevD.23.287} {\bibfield  {journal}
  {\bibinfo  {journal} {Phys. Rev. D}\ }\textbf {\bibinfo {volume} {23}},\
  \bibinfo {pages} {287} (\bibinfo {year} {1981})}\BibitemShut {NoStop}%
\bibitem [{\citenamefont {Hawking}(1975)}]{Hawking:1975vcx}%
  \BibitemOpen
  \bibfield  {author} {\bibinfo {author} {\bibfnamefont {S.~W.}\ \bibnamefont
  {Hawking}},\ }\bibfield  {title} {\enquote {\bibinfo {title} {{Particle
  Creation by Black Holes}},}\ }\href {\doibase 10.1007/BF02345020} {\bibfield
  {journal} {\bibinfo  {journal} {Commun. Math. Phys.}\ }\textbf {\bibinfo
  {volume} {43}},\ \bibinfo {pages} {199--220} (\bibinfo {year} {1975})},\
  \bibinfo {note} {[Erratum: Commun.Math.Phys. 46, 206 (1976)]}\BibitemShut
  {NoStop}%
\bibitem [{\citenamefont {Dvali}(2018{\natexlab{a}})}]{Dvali:2018xpy}%
  \BibitemOpen
  \bibfield  {author} {\bibinfo {author} {\bibfnamefont {Gia}\ \bibnamefont
  {Dvali}},\ }\bibfield  {title} {\enquote {\bibinfo {title} {{A Microscopic
  Model of Holography: Survival by the Burden of Memory}},}\ }\href@noop {} {\
  (\bibinfo {year} {2018}{\natexlab{a}})},\ \Eprint
  {http://arxiv.org/abs/1810.02336} {arXiv:1810.02336 [hep-th]} \BibitemShut
  {NoStop}%
\bibitem [{\citenamefont {Dvali}(2021{\natexlab{a}})}]{Dvali:2020wqi}%
  \BibitemOpen
  \bibfield  {author} {\bibinfo {author} {\bibfnamefont {Gia}\ \bibnamefont
  {Dvali}},\ }\bibfield  {title} {\enquote {\bibinfo {title} {{Entropy Bound
  and Unitarity of Scattering Amplitudes}},}\ }\href {\doibase
  10.1007/JHEP03(2021)126} {\bibfield  {journal} {\bibinfo  {journal} {JHEP}\
  }\textbf {\bibinfo {volume} {03}},\ \bibinfo {pages} {126} (\bibinfo {year}
  {2021}{\natexlab{a}})},\ \Eprint {http://arxiv.org/abs/2003.05546}
  {arXiv:2003.05546 [hep-th]} \BibitemShut {NoStop}%
\bibitem [{\citenamefont {Dvali}(2021{\natexlab{b}})}]{Dvali:2019jjw}%
  \BibitemOpen
  \bibfield  {author} {\bibinfo {author} {\bibfnamefont {Gia}\ \bibnamefont
  {Dvali}},\ }\bibfield  {title} {\enquote {\bibinfo {title} {{Area Law
  Saturation of Entropy Bound from Perturbative Unitarity in Renormalizable
  Theories}},}\ }\href {\doibase 10.1002/prop.202000090} {\bibfield  {journal}
  {\bibinfo  {journal} {Fortsch. Phys.}\ }\textbf {\bibinfo {volume} {69}},\
  \bibinfo {pages} {2000090} (\bibinfo {year} {2021}{\natexlab{b}})},\ \Eprint
  {http://arxiv.org/abs/1906.03530} {arXiv:1906.03530 [hep-th]} \BibitemShut
  {NoStop}%
\bibitem [{\citenamefont {Dvali}(2021{\natexlab{c}})}]{Dvali:2019ulr}%
  \BibitemOpen
  \bibfield  {author} {\bibinfo {author} {\bibfnamefont {Gia}\ \bibnamefont
  {Dvali}},\ }\bibfield  {title} {\enquote {\bibinfo {title} {{Unitarity
  Entropy Bound: Solitons and Instantons}},}\ }\href {\doibase
  10.1002/prop.202000091} {\bibfield  {journal} {\bibinfo  {journal} {Fortsch.
  Phys.}\ }\textbf {\bibinfo {volume} {69}},\ \bibinfo {pages} {2000091}
  (\bibinfo {year} {2021}{\natexlab{c}})},\ \Eprint
  {http://arxiv.org/abs/1907.07332} {arXiv:1907.07332 [hep-th]} \BibitemShut
  {NoStop}%
\bibitem [{\citenamefont {Dvali}\ and\ \citenamefont
  {Gomez}(2013{\natexlab{a}})}]{Dvali:2011aa}%
  \BibitemOpen
  \bibfield  {author} {\bibinfo {author} {\bibfnamefont {Gia}\ \bibnamefont
  {Dvali}}\ and\ \bibinfo {author} {\bibfnamefont {Cesar}\ \bibnamefont
  {Gomez}},\ }\bibfield  {title} {\enquote {\bibinfo {title} {{Black Hole's
  Quantum N-Portrait}},}\ }\href {\doibase 10.1002/prop.201300001} {\bibfield
  {journal} {\bibinfo  {journal} {Fortsch. Phys.}\ }\textbf {\bibinfo {volume}
  {61}},\ \bibinfo {pages} {742--767} (\bibinfo {year} {2013}{\natexlab{a}})},\
  \Eprint {http://arxiv.org/abs/1112.3359} {arXiv:1112.3359 [hep-th]}
  \BibitemShut {NoStop}%
\bibitem [{\citenamefont {Dvali}(2021{\natexlab{d}})}]{Dvali:2021jto}%
  \BibitemOpen
  \bibfield  {author} {\bibinfo {author} {\bibfnamefont {Gia}\ \bibnamefont
  {Dvali}},\ }\bibfield  {title} {\enquote {\bibinfo {title} {{Bounds on
  quantum information storage and retrieval}},}\ }\href {\doibase
  10.1098/rsta.2021.0071} {\bibfield  {journal} {\bibinfo  {journal} {Phil.
  Trans. A. Math. Phys. Eng. Sci.}\ }\textbf {\bibinfo {volume} {380}},\
  \bibinfo {pages} {20210071} (\bibinfo {year} {2021}{\natexlab{d}})},\ \Eprint
  {http://arxiv.org/abs/2107.10616} {arXiv:2107.10616 [hep-th]} \BibitemShut
  {NoStop}%
\bibitem [{\citenamefont {Dvali}\ and\ \citenamefont
  {Venugopalan}(2022)}]{Dvali:2021ooc}%
  \BibitemOpen
  \bibfield  {author} {\bibinfo {author} {\bibfnamefont {Gia}\ \bibnamefont
  {Dvali}}\ and\ \bibinfo {author} {\bibfnamefont {Raju}\ \bibnamefont
  {Venugopalan}},\ }\bibfield  {title} {\enquote {\bibinfo {title}
  {{Classicalization and unitarization of wee partons in QCD and gravity: The
  CGC-black hole correspondence}},}\ }\href {\doibase
  10.1103/PhysRevD.105.056026} {\bibfield  {journal} {\bibinfo  {journal}
  {Phys. Rev. D}\ }\textbf {\bibinfo {volume} {105}},\ \bibinfo {pages}
  {056026} (\bibinfo {year} {2022})},\ \Eprint
  {http://arxiv.org/abs/2106.11989} {arXiv:2106.11989 [hep-th]} \BibitemShut
  {NoStop}%
\bibitem [{\citenamefont {Dvali}\ and\ \citenamefont
  {Sakhelashvili}(2022)}]{Dvali:2021rlf}%
  \BibitemOpen
  \bibfield  {author} {\bibinfo {author} {\bibfnamefont {Gia}\ \bibnamefont
  {Dvali}}\ and\ \bibinfo {author} {\bibfnamefont {Otari}\ \bibnamefont
  {Sakhelashvili}},\ }\bibfield  {title} {\enquote {\bibinfo {title}
  {{Black-hole-like saturons in Gross-Neveu}},}\ }\href {\doibase
  10.1103/PhysRevD.105.065014} {\bibfield  {journal} {\bibinfo  {journal}
  {Phys. Rev. D}\ }\textbf {\bibinfo {volume} {105}},\ \bibinfo {pages}
  {065014} (\bibinfo {year} {2022})},\ \Eprint
  {http://arxiv.org/abs/2111.03620} {arXiv:2111.03620 [hep-th]} \BibitemShut
  {NoStop}%
\bibitem [{\citenamefont {Dvali}\ \emph
  {et~al.}(2022{\natexlab{a}})\citenamefont {Dvali}, \citenamefont {Kaikov},\
  and\ \citenamefont {Berm\'udez}}]{Dvali:2021tez}%
  \BibitemOpen
  \bibfield  {author} {\bibinfo {author} {\bibfnamefont {Gia}\ \bibnamefont
  {Dvali}}, \bibinfo {author} {\bibfnamefont {Oleg}\ \bibnamefont {Kaikov}}, \
  and\ \bibinfo {author} {\bibfnamefont {Juan Sebasti\'an~Valbuena}\
  \bibnamefont {Berm\'udez}},\ }\bibfield  {title} {\enquote {\bibinfo {title}
  {{How special are black holes? Correspondence with objects saturating
  unitarity bounds in generic theories}},}\ }\href {\doibase
  10.1103/PhysRevD.105.056013} {\bibfield  {journal} {\bibinfo  {journal}
  {Phys. Rev. D}\ }\textbf {\bibinfo {volume} {105}},\ \bibinfo {pages}
  {056013} (\bibinfo {year} {2022}{\natexlab{a}})},\ \Eprint
  {http://arxiv.org/abs/2112.00551} {arXiv:2112.00551 [hep-th]} \BibitemShut
  {NoStop}%
\bibitem [{\citenamefont {Dvali}\ \emph
  {et~al.}(2022{\natexlab{b}})\citenamefont {Dvali}, \citenamefont
  {K{\"u}hnel},\ and\ \citenamefont {Zantedeschi}}]{Dvali:2021ofp}%
  \BibitemOpen
  \bibfield  {author} {\bibinfo {author} {\bibfnamefont {Gia}\ \bibnamefont
  {Dvali}}, \bibinfo {author} {\bibfnamefont {Florian}\ \bibnamefont
  {K{\"u}hnel}}, \ and\ \bibinfo {author} {\bibfnamefont {Michael}\
  \bibnamefont {Zantedeschi}},\ }\bibfield  {title} {\enquote {\bibinfo {title}
  {{Vortices in Black Holes}},}\ }\href {\doibase
  10.1103/PhysRevLett.129.061302} {\bibfield  {journal} {\bibinfo  {journal}
  {Phys. Rev. Lett.}\ }\textbf {\bibinfo {volume} {129}},\ \bibinfo {pages}
  {061302} (\bibinfo {year} {2022}{\natexlab{b}})},\ \Eprint
  {http://arxiv.org/abs/2112.08354} {arXiv:2112.08354 [hep-th]} \BibitemShut
  {NoStop}%
\bibitem [{\citenamefont {Dvali}\ \emph
  {et~al.}(2024{\natexlab{a}})\citenamefont {Dvali}, \citenamefont {Kaikov},
  \citenamefont {K{\"u}hnel}, \citenamefont {Valbuena-Bermudez},\ and\
  \citenamefont {Zantedeschi}}]{Dvali:2023qlk}%
  \BibitemOpen
  \bibfield  {author} {\bibinfo {author} {\bibfnamefont {Gia}\ \bibnamefont
  {Dvali}}, \bibinfo {author} {\bibfnamefont {Oleg}\ \bibnamefont {Kaikov}},
  \bibinfo {author} {\bibfnamefont {Florian}\ \bibnamefont {K{\"u}hnel}},
  \bibinfo {author} {\bibfnamefont {Juan~Sebastian}\ \bibnamefont
  {Valbuena-Bermudez}}, \ and\ \bibinfo {author} {\bibfnamefont {Michael}\
  \bibnamefont {Zantedeschi}},\ }\bibfield  {title} {\enquote {\bibinfo {title}
  {{Vortex Effects in Merging Black Holes and Saturons}},}\ }\href {\doibase
  10.1103/PhysRevLett.132.151402} {\bibfield  {journal} {\bibinfo  {journal}
  {Phys. Rev. Lett.}\ }\textbf {\bibinfo {volume} {132}},\ \bibinfo {pages}
  {151402} (\bibinfo {year} {2024}{\natexlab{a}})},\ \Eprint
  {http://arxiv.org/abs/2310.02288} {arXiv:2310.02288 [hep-ph]} \BibitemShut
  {NoStop}%
\bibitem [{\citenamefont {Dvali}\ \emph {et~al.}(2020)\citenamefont {Dvali},
  \citenamefont {Eisemann}, \citenamefont {Michel},\ and\ \citenamefont
  {Zell}}]{Dvali:2020wft}%
  \BibitemOpen
  \bibfield  {author} {\bibinfo {author} {\bibfnamefont {Gia}\ \bibnamefont
  {Dvali}}, \bibinfo {author} {\bibfnamefont {Lukas}\ \bibnamefont {Eisemann}},
  \bibinfo {author} {\bibfnamefont {Marco}\ \bibnamefont {Michel}}, \ and\
  \bibinfo {author} {\bibfnamefont {Sebastian}\ \bibnamefont {Zell}},\
  }\bibfield  {title} {\enquote {\bibinfo {title} {{Black hole metamorphosis
  and stabilization by memory burden}},}\ }\href {\doibase
  10.1103/PhysRevD.102.103523} {\bibfield  {journal} {\bibinfo  {journal}
  {Phys. Rev. D}\ }\textbf {\bibinfo {volume} {102}},\ \bibinfo {pages}
  {103523} (\bibinfo {year} {2020})},\ \Eprint
  {http://arxiv.org/abs/2006.00011} {arXiv:2006.00011 [hep-th]} \BibitemShut
  {NoStop}%
\bibitem [{\citenamefont {Dvali}\ \emph {et~al.}(2019)\citenamefont {Dvali},
  \citenamefont {Eisemann}, \citenamefont {Michel},\ and\ \citenamefont
  {Zell}}]{Dvali:2018ytn}%
  \BibitemOpen
  \bibfield  {author} {\bibinfo {author} {\bibfnamefont {Gia}\ \bibnamefont
  {Dvali}}, \bibinfo {author} {\bibfnamefont {Lukas}\ \bibnamefont {Eisemann}},
  \bibinfo {author} {\bibfnamefont {Marco}\ \bibnamefont {Michel}}, \ and\
  \bibinfo {author} {\bibfnamefont {Sebastian}\ \bibnamefont {Zell}},\
  }\bibfield  {title} {\enquote {\bibinfo {title} {{Universe's Primordial
  Quantum Memories}},}\ }\href {\doibase 10.1088/1475-7516/2019/03/010}
  {\bibfield  {journal} {\bibinfo  {journal} {JCAP}\ }\textbf {\bibinfo
  {volume} {03}},\ \bibinfo {pages} {010} (\bibinfo {year} {2019})},\ \Eprint
  {http://arxiv.org/abs/1812.08749} {arXiv:1812.08749 [hep-th]} \BibitemShut
  {NoStop}%
\bibitem [{\citenamefont {Dvali}\ \emph
  {et~al.}(2024{\natexlab{b}})\citenamefont {Dvali}, \citenamefont
  {Valbuena-Berm\'udez},\ and\ \citenamefont {Zantedeschi}}]{Dvali:2024hsb}%
  \BibitemOpen
  \bibfield  {author} {\bibinfo {author} {\bibfnamefont {Gia}\ \bibnamefont
  {Dvali}}, \bibinfo {author} {\bibfnamefont {Juan~Sebasti\'an}\ \bibnamefont
  {Valbuena-Berm\'udez}}, \ and\ \bibinfo {author} {\bibfnamefont {Michael}\
  \bibnamefont {Zantedeschi}},\ }\bibfield  {title} {\enquote {\bibinfo {title}
  {{Memory burden effect in black holes and solitons: Implications for PBH}},}\
  }\href {\doibase 10.1103/PhysRevD.110.056029} {\bibfield  {journal} {\bibinfo
   {journal} {Phys. Rev. D}\ }\textbf {\bibinfo {volume} {110}},\ \bibinfo
  {pages} {056029} (\bibinfo {year} {2024}{\natexlab{b}})},\ \Eprint
  {http://arxiv.org/abs/2405.13117} {arXiv:2405.13117 [hep-th]} \BibitemShut
  {NoStop}%
\bibitem [{\citenamefont {Alexandre}\ \emph {et~al.}(2024)\citenamefont
  {Alexandre}, \citenamefont {Dvali},\ and\ \citenamefont
  {Koutsangelas}}]{Alexandre:2024nuo}%
  \BibitemOpen
  \bibfield  {author} {\bibinfo {author} {\bibfnamefont {Ana}\ \bibnamefont
  {Alexandre}}, \bibinfo {author} {\bibfnamefont {Gia}\ \bibnamefont {Dvali}},
  \ and\ \bibinfo {author} {\bibfnamefont {Emmanouil}\ \bibnamefont
  {Koutsangelas}},\ }\bibfield  {title} {\enquote {\bibinfo {title} {{New mass
  window for primordial black holes as dark matter from the memory burden
  effect}},}\ }\href {\doibase 10.1103/PhysRevD.110.036004} {\bibfield
  {journal} {\bibinfo  {journal} {Phys. Rev. D}\ }\textbf {\bibinfo {volume}
  {110}},\ \bibinfo {pages} {036004} (\bibinfo {year} {2024})},\ \Eprint
  {http://arxiv.org/abs/2402.14069} {arXiv:2402.14069 [hep-ph]} \BibitemShut
  {NoStop}%
\bibitem [{\citenamefont {Thoss}\ \emph {et~al.}(2024)\citenamefont {Thoss},
  \citenamefont {Burkert},\ and\ \citenamefont {Kohri}}]{Thoss:2024hsr}%
  \BibitemOpen
  \bibfield  {author} {\bibinfo {author} {\bibfnamefont {Valentin}\
  \bibnamefont {Thoss}}, \bibinfo {author} {\bibfnamefont {Andreas}\
  \bibnamefont {Burkert}}, \ and\ \bibinfo {author} {\bibfnamefont {Kazunori}\
  \bibnamefont {Kohri}},\ }\bibfield  {title} {\enquote {\bibinfo {title}
  {{Breakdown of hawking evaporation opens new mass window for primordial black
  holes as dark matter candidate}},}\ }\href {\doibase 10.1093/mnras/stae1098}
  {\bibfield  {journal} {\bibinfo  {journal} {Mon. Not. Roy. Astron. Soc.}\
  }\textbf {\bibinfo {volume} {532}},\ \bibinfo {pages} {451--459} (\bibinfo
  {year} {2024})},\ \Eprint {http://arxiv.org/abs/2402.17823} {arXiv:2402.17823
  [astro-ph.CO]} \BibitemShut {NoStop}%
\bibitem [{\citenamefont {Dvali}\ and\ \citenamefont
  {Gomez}(2014{\natexlab{a}})}]{Dvali:2012en}%
  \BibitemOpen
  \bibfield  {author} {\bibinfo {author} {\bibfnamefont {Gia}\ \bibnamefont
  {Dvali}}\ and\ \bibinfo {author} {\bibfnamefont {Cesar}\ \bibnamefont
  {Gomez}},\ }\bibfield  {title} {\enquote {\bibinfo {title} {{Black Holes as
  Critical Point of Quantum Phase Transition}},}\ }\href {\doibase
  10.1140/epjc/s10052-014-2752-3} {\bibfield  {journal} {\bibinfo  {journal}
  {Eur. Phys. J. C}\ }\textbf {\bibinfo {volume} {74}},\ \bibinfo {pages}
  {2752} (\bibinfo {year} {2014}{\natexlab{a}})},\ \Eprint
  {http://arxiv.org/abs/1207.4059} {arXiv:1207.4059 [hep-th]} \BibitemShut
  {NoStop}%
\bibitem [{\citenamefont {Dvali}\ and\ \citenamefont
  {Gomez}(2014{\natexlab{b}})}]{Dvali:2013eja}%
  \BibitemOpen
  \bibfield  {author} {\bibinfo {author} {\bibfnamefont {Gia}\ \bibnamefont
  {Dvali}}\ and\ \bibinfo {author} {\bibfnamefont {Cesar}\ \bibnamefont
  {Gomez}},\ }\bibfield  {title} {\enquote {\bibinfo {title} {{Quantum
  Compositeness of Gravity: Black Holes, AdS and Inflation}},}\ }\href
  {\doibase 10.1088/1475-7516/2014/01/023} {\bibfield  {journal} {\bibinfo
  {journal} {JCAP}\ }\textbf {\bibinfo {volume} {01}},\ \bibinfo {pages} {023}
  (\bibinfo {year} {2014}{\natexlab{b}})},\ \Eprint
  {http://arxiv.org/abs/1312.4795} {arXiv:1312.4795 [hep-th]} \BibitemShut
  {NoStop}%
\bibitem [{\citenamefont {Dvali}\ and\ \citenamefont
  {Gomez}(2013{\natexlab{b}})}]{Dvali:2012rt}%
  \BibitemOpen
  \bibfield  {author} {\bibinfo {author} {\bibfnamefont {Gia}\ \bibnamefont
  {Dvali}}\ and\ \bibinfo {author} {\bibfnamefont {Cesar}\ \bibnamefont
  {Gomez}},\ }\bibfield  {title} {\enquote {\bibinfo {title} {{Black Hole's 1/N
  Hair}},}\ }\href {\doibase 10.1016/j.physletb.2013.01.020} {\bibfield
  {journal} {\bibinfo  {journal} {Phys. Lett. B}\ }\textbf {\bibinfo {volume}
  {719}},\ \bibinfo {pages} {419--423} (\bibinfo {year}
  {2013}{\natexlab{b}})},\ \Eprint {http://arxiv.org/abs/1203.6575}
  {arXiv:1203.6575 [hep-th]} \BibitemShut {NoStop}%
\bibitem [{\citenamefont {Dvali}\ and\ \citenamefont
  {Gomez}(2012)}]{Dvali:2012wq}%
  \BibitemOpen
  \bibfield  {author} {\bibinfo {author} {\bibfnamefont {Gia}\ \bibnamefont
  {Dvali}}\ and\ \bibinfo {author} {\bibfnamefont {Cesar}\ \bibnamefont
  {Gomez}},\ }\bibfield  {title} {\enquote {\bibinfo {title} {{Black Hole
  Macro-Quantumness}},}\ }\href@noop {} {\  (\bibinfo {year} {2012})},\ \Eprint
  {http://arxiv.org/abs/1212.0765} {arXiv:1212.0765 [hep-th]} \BibitemShut
  {NoStop}%
\bibitem [{\citenamefont {Dvali}\ \emph {et~al.}(2013)\citenamefont {Dvali},
  \citenamefont {Flassig}, \citenamefont {Gomez}, \citenamefont {Pritzel},\
  and\ \citenamefont {Wintergerst}}]{Dvali:2013vxa}%
  \BibitemOpen
  \bibfield  {author} {\bibinfo {author} {\bibfnamefont {Gia}\ \bibnamefont
  {Dvali}}, \bibinfo {author} {\bibfnamefont {Daniel}\ \bibnamefont {Flassig}},
  \bibinfo {author} {\bibfnamefont {Cesar}\ \bibnamefont {Gomez}}, \bibinfo
  {author} {\bibfnamefont {Alexander}\ \bibnamefont {Pritzel}}, \ and\ \bibinfo
  {author} {\bibfnamefont {Nico}\ \bibnamefont {Wintergerst}},\ }\bibfield
  {title} {\enquote {\bibinfo {title} {{Scrambling in the Black Hole
  Portrait}},}\ }\href {\doibase 10.1103/PhysRevD.88.124041} {\bibfield
  {journal} {\bibinfo  {journal} {Phys. Rev. D}\ }\textbf {\bibinfo {volume}
  {88}},\ \bibinfo {pages} {124041} (\bibinfo {year} {2013})},\ \Eprint
  {http://arxiv.org/abs/1307.3458} {arXiv:1307.3458 [hep-th]} \BibitemShut
  {NoStop}%
\bibitem [{\citenamefont {Dvali}\ \emph {et~al.}(2017)\citenamefont {Dvali},
  \citenamefont {Gomez},\ and\ \citenamefont {Zell}}]{Dvali:2017eba}%
  \BibitemOpen
  \bibfield  {author} {\bibinfo {author} {\bibfnamefont {Gia}\ \bibnamefont
  {Dvali}}, \bibinfo {author} {\bibfnamefont {Cesar}\ \bibnamefont {Gomez}}, \
  and\ \bibinfo {author} {\bibfnamefont {Sebastian}\ \bibnamefont {Zell}},\
  }\bibfield  {title} {\enquote {\bibinfo {title} {{Quantum Break-Time of de
  Sitter}},}\ }\href {\doibase 10.1088/1475-7516/2017/06/028} {\bibfield
  {journal} {\bibinfo  {journal} {JCAP}\ }\textbf {\bibinfo {volume} {06}},\
  \bibinfo {pages} {028} (\bibinfo {year} {2017})},\ \Eprint
  {http://arxiv.org/abs/1701.08776} {arXiv:1701.08776 [hep-th]} \BibitemShut
  {NoStop}%
\bibitem [{\citenamefont {Dvali}\ and\ \citenamefont
  {Gomez}(2016)}]{Dvali:2014gua}%
  \BibitemOpen
  \bibfield  {author} {\bibinfo {author} {\bibfnamefont {Gia}\ \bibnamefont
  {Dvali}}\ and\ \bibinfo {author} {\bibfnamefont {Cesar}\ \bibnamefont
  {Gomez}},\ }\bibfield  {title} {\enquote {\bibinfo {title} {{Quantum
  Exclusion of Positive Cosmological Constant?}}}\ }\href {\doibase
  10.1002/andp.201500216} {\bibfield  {journal} {\bibinfo  {journal} {Annalen
  Phys.}\ }\textbf {\bibinfo {volume} {528}},\ \bibinfo {pages} {68--73}
  (\bibinfo {year} {2016})},\ \Eprint {http://arxiv.org/abs/1412.8077}
  {arXiv:1412.8077 [hep-th]} \BibitemShut {NoStop}%
\bibitem [{\citenamefont {Dvali}(2025)}]{Dvali:2024dzf}%
  \BibitemOpen
  \bibfield  {author} {\bibinfo {author} {\bibfnamefont {Gia}\ \bibnamefont
  {Dvali}},\ }\bibfield  {title} {\enquote {\bibinfo {title} {{A string
  theoretic derivation of gibbons-hawking entropy}},}\ }\href {\doibase
  10.1007/s10714-025-03446-6} {\bibfield  {journal} {\bibinfo  {journal} {Gen.
  Rel. Grav.}\ }\textbf {\bibinfo {volume} {57}},\ \bibinfo {pages} {118}
  (\bibinfo {year} {2025})},\ \Eprint {http://arxiv.org/abs/2407.01510}
  {arXiv:2407.01510 [hep-th]} \BibitemShut {NoStop}%
\bibitem [{\citenamefont {Dvali}\ and\ \citenamefont
  {Eisemann}(2022)}]{Dvali:2022vzz}%
  \BibitemOpen
  \bibfield  {author} {\bibinfo {author} {\bibfnamefont {Gia}\ \bibnamefont
  {Dvali}}\ and\ \bibinfo {author} {\bibfnamefont {Lukas}\ \bibnamefont
  {Eisemann}},\ }\bibfield  {title} {\enquote {\bibinfo {title} {{Perturbative
  understanding of nonperturbative processes and quantumization versus
  classicalization}},}\ }\href {\doibase 10.1103/PhysRevD.106.125019}
  {\bibfield  {journal} {\bibinfo  {journal} {Phys. Rev. D}\ }\textbf {\bibinfo
  {volume} {106}},\ \bibinfo {pages} {125019} (\bibinfo {year} {2022})},\
  \Eprint {http://arxiv.org/abs/2211.02618} {arXiv:2211.02618 [hep-th]}
  \BibitemShut {NoStop}%
\bibitem [{\citenamefont {Berezhiani}\ and\ \citenamefont
  {Zantedeschi}(2021)}]{Berezhiani:2020pbv}%
  \BibitemOpen
  \bibfield  {author} {\bibinfo {author} {\bibfnamefont {Lasha}\ \bibnamefont
  {Berezhiani}}\ and\ \bibinfo {author} {\bibfnamefont {Michael}\ \bibnamefont
  {Zantedeschi}},\ }\bibfield  {title} {\enquote {\bibinfo {title} {{Evolution
  of coherent states as quantum counterpart of classical dynamics}},}\ }\href
  {\doibase 10.1103/PhysRevD.104.085007} {\bibfield  {journal} {\bibinfo
  {journal} {Phys. Rev. D}\ }\textbf {\bibinfo {volume} {104}},\ \bibinfo
  {pages} {085007} (\bibinfo {year} {2021})},\ \Eprint
  {http://arxiv.org/abs/2011.11229} {arXiv:2011.11229 [hep-th]} \BibitemShut
  {NoStop}%
\bibitem [{\citenamefont {Berezhiani}\ \emph
  {et~al.}(2022{\natexlab{a}})\citenamefont {Berezhiani}, \citenamefont
  {Cintia},\ and\ \citenamefont {Zantedeschi}}]{Berezhiani:2021gph}%
  \BibitemOpen
  \bibfield  {author} {\bibinfo {author} {\bibfnamefont {Lasha}\ \bibnamefont
  {Berezhiani}}, \bibinfo {author} {\bibfnamefont {Giordano}\ \bibnamefont
  {Cintia}}, \ and\ \bibinfo {author} {\bibfnamefont {Michael}\ \bibnamefont
  {Zantedeschi}},\ }\bibfield  {title} {\enquote {\bibinfo {title}
  {{Background-field method and initial-time singularity for coherent
  states}},}\ }\href {\doibase 10.1103/PhysRevD.105.045003} {\bibfield
  {journal} {\bibinfo  {journal} {Phys. Rev. D}\ }\textbf {\bibinfo {volume}
  {105}},\ \bibinfo {pages} {045003} (\bibinfo {year} {2022}{\natexlab{a}})},\
  \Eprint {http://arxiv.org/abs/2108.13235} {arXiv:2108.13235 [hep-th]}
  \BibitemShut {NoStop}%
\bibitem [{\citenamefont {Berezhiani}\ \emph
  {et~al.}(2022{\natexlab{b}})\citenamefont {Berezhiani}, \citenamefont
  {Dvali},\ and\ \citenamefont {Sakhelashvili}}]{Berezhiani:2021zst}%
  \BibitemOpen
  \bibfield  {author} {\bibinfo {author} {\bibfnamefont {Lasha}\ \bibnamefont
  {Berezhiani}}, \bibinfo {author} {\bibfnamefont {Gia}\ \bibnamefont {Dvali}},
  \ and\ \bibinfo {author} {\bibfnamefont {Otari}\ \bibnamefont
  {Sakhelashvili}},\ }\bibfield  {title} {\enquote {\bibinfo {title} {{de
  Sitter space as a BRST invariant coherent state of gravitons}},}\ }\href
  {\doibase 10.1103/PhysRevD.105.025022} {\bibfield  {journal} {\bibinfo
  {journal} {Phys. Rev. D}\ }\textbf {\bibinfo {volume} {105}},\ \bibinfo
  {pages} {025022} (\bibinfo {year} {2022}{\natexlab{b}})},\ \Eprint
  {http://arxiv.org/abs/2111.12022} {arXiv:2111.12022 [hep-th]} \BibitemShut
  {NoStop}%
\bibitem [{\citenamefont {Berezhiani}\ \emph {et~al.}(2024)\citenamefont
  {Berezhiani}, \citenamefont {Cintia},\ and\ \citenamefont
  {Zantedeschi}}]{Berezhiani:2023uwt}%
  \BibitemOpen
  \bibfield  {author} {\bibinfo {author} {\bibfnamefont {Lasha}\ \bibnamefont
  {Berezhiani}}, \bibinfo {author} {\bibfnamefont {Giordano}\ \bibnamefont
  {Cintia}}, \ and\ \bibinfo {author} {\bibfnamefont {Michael}\ \bibnamefont
  {Zantedeschi}},\ }\bibfield  {title} {\enquote {\bibinfo {title}
  {{Perturbative construction of coherent states}},}\ }\href {\doibase
  10.1103/PhysRevD.109.085018} {\bibfield  {journal} {\bibinfo  {journal}
  {Phys. Rev. D}\ }\textbf {\bibinfo {volume} {109}},\ \bibinfo {pages}
  {085018} (\bibinfo {year} {2024})},\ \Eprint
  {http://arxiv.org/abs/2311.18650} {arXiv:2311.18650 [hep-th]} \BibitemShut
  {NoStop}%
\bibitem [{\citenamefont {Berezhiani}\ \emph {et~al.}(2025)\citenamefont
  {Berezhiani}, \citenamefont {Dvali},\ and\ \citenamefont
  {Sakhelashvili}}]{Berezhiani:2024pub}%
  \BibitemOpen
  \bibfield  {author} {\bibinfo {author} {\bibfnamefont {Lasha}\ \bibnamefont
  {Berezhiani}}, \bibinfo {author} {\bibfnamefont {Gia}\ \bibnamefont {Dvali}},
  \ and\ \bibinfo {author} {\bibfnamefont {Otari}\ \bibnamefont
  {Sakhelashvili}},\ }\bibfield  {title} {\enquote {\bibinfo {title} {{Coherent
  states in gauge theories: Topological defects and other classical
  configurations}},}\ }\href {\doibase 10.1103/PhysRevD.111.065018} {\bibfield
  {journal} {\bibinfo  {journal} {Phys. Rev. D}\ }\textbf {\bibinfo {volume}
  {111}},\ \bibinfo {pages} {065018} (\bibinfo {year} {2025})},\ \Eprint
  {http://arxiv.org/abs/2411.11657} {arXiv:2411.11657 [hep-th]} \BibitemShut
  {NoStop}%
\bibitem [{\citenamefont {Dvali}\ and\ \citenamefont
  {Zell}(2018)}]{Dvali:2017ruz}%
  \BibitemOpen
  \bibfield  {author} {\bibinfo {author} {\bibfnamefont {Gia}\ \bibnamefont
  {Dvali}}\ and\ \bibinfo {author} {\bibfnamefont {Sebastian}\ \bibnamefont
  {Zell}},\ }\bibfield  {title} {\enquote {\bibinfo {title} {{Classicality and
  Quantum Break-Time for Cosmic Axions}},}\ }\href {\doibase
  10.1088/1475-7516/2018/07/064} {\bibfield  {journal} {\bibinfo  {journal}
  {JCAP}\ }\textbf {\bibinfo {volume} {07}},\ \bibinfo {pages} {064} (\bibinfo
  {year} {2018})},\ \Eprint {http://arxiv.org/abs/1710.00835} {arXiv:1710.00835
  [hep-ph]} \BibitemShut {NoStop}%
\bibitem [{\citenamefont {Gross}\ and\ \citenamefont
  {Neveu}(1974)}]{Gross:1974jv}%
  \BibitemOpen
  \bibfield  {author} {\bibinfo {author} {\bibfnamefont {David~J.}\
  \bibnamefont {Gross}}\ and\ \bibinfo {author} {\bibfnamefont {Andre}\
  \bibnamefont {Neveu}},\ }\bibfield  {title} {\enquote {\bibinfo {title}
  {{Dynamical Symmetry Breaking in Asymptotically Free Field Theories}},}\
  }\href {\doibase 10.1103/PhysRevD.10.3235} {\bibfield  {journal} {\bibinfo
  {journal} {Phys. Rev. D}\ }\textbf {\bibinfo {volume} {10}},\ \bibinfo
  {pages} {3235} (\bibinfo {year} {1974})}\BibitemShut {NoStop}%
\bibitem [{\citenamefont {Dvali}\ and\ \citenamefont
  {Vilenkin}(2003)}]{Dvali:2002fi}%
  \BibitemOpen
  \bibfield  {author} {\bibinfo {author} {\bibfnamefont {Gia}\ \bibnamefont
  {Dvali}}\ and\ \bibinfo {author} {\bibfnamefont {Alexander}\ \bibnamefont
  {Vilenkin}},\ }\bibfield  {title} {\enquote {\bibinfo {title} {{Solitonic
  D-branes and brane annihilation}},}\ }\href {\doibase
  10.1103/PhysRevD.67.046002} {\bibfield  {journal} {\bibinfo  {journal} {Phys.
  Rev. D}\ }\textbf {\bibinfo {volume} {67}},\ \bibinfo {pages} {046002}
  (\bibinfo {year} {2003})},\ \Eprint {http://arxiv.org/abs/hep-th/0209217}
  {arXiv:hep-th/0209217} \BibitemShut {NoStop}%
\bibitem [{\citenamefont {Dvali}\ \emph
  {et~al.}(2015{\natexlab{a}})\citenamefont {Dvali}, \citenamefont {Franca},
  \citenamefont {Gomez},\ and\ \citenamefont {Wintergerst}}]{Dvali:2015ywa}%
  \BibitemOpen
  \bibfield  {author} {\bibinfo {author} {\bibfnamefont {Gia}\ \bibnamefont
  {Dvali}}, \bibinfo {author} {\bibfnamefont {Andre}\ \bibnamefont {Franca}},
  \bibinfo {author} {\bibfnamefont {Cesar}\ \bibnamefont {Gomez}}, \ and\
  \bibinfo {author} {\bibfnamefont {Nico}\ \bibnamefont {Wintergerst}},\
  }\bibfield  {title} {\enquote {\bibinfo {title} {{Nambu-Goldstone Effective
  Theory of Information at Quantum Criticality}},}\ }\href {\doibase
  10.1103/PhysRevD.92.125002} {\bibfield  {journal} {\bibinfo  {journal} {Phys.
  Rev. D}\ }\textbf {\bibinfo {volume} {92}},\ \bibinfo {pages} {125002}
  (\bibinfo {year} {2015}{\natexlab{a}})},\ \Eprint
  {http://arxiv.org/abs/1507.02948} {arXiv:1507.02948 [hep-th]} \BibitemShut
  {NoStop}%
\bibitem [{\citenamefont {Lee}\ and\ \citenamefont {Wick}(1974)}]{Lee:1974ma}%
  \BibitemOpen
  \bibfield  {author} {\bibinfo {author} {\bibfnamefont {T.~D.}\ \bibnamefont
  {Lee}}\ and\ \bibinfo {author} {\bibfnamefont {G.~C.}\ \bibnamefont {Wick}},\
  }\bibfield  {title} {\enquote {\bibinfo {title} {{Vacuum Stability and Vacuum
  Excitation in a Spin 0 Field Theory}},}\ }\href {\doibase
  10.1103/PhysRevD.9.2291} {\bibfield  {journal} {\bibinfo  {journal} {Phys.
  Rev. D}\ }\textbf {\bibinfo {volume} {9}},\ \bibinfo {pages} {2291--2316}
  (\bibinfo {year} {1974})}\BibitemShut {NoStop}%
\bibitem [{\citenamefont {Friedberg}\ \emph {et~al.}(1976)\citenamefont
  {Friedberg}, \citenamefont {Lee},\ and\ \citenamefont
  {Sirlin}}]{Friedberg:1976me}%
  \BibitemOpen
  \bibfield  {author} {\bibinfo {author} {\bibfnamefont {R.}~\bibnamefont
  {Friedberg}}, \bibinfo {author} {\bibfnamefont {T.~D.}\ \bibnamefont {Lee}},
  \ and\ \bibinfo {author} {\bibfnamefont {A.}~\bibnamefont {Sirlin}},\
  }\bibfield  {title} {\enquote {\bibinfo {title} {{A Class of Scalar-Field
  Soliton Solutions in Three Space Dimensions}},}\ }\href {\doibase
  10.1103/PhysRevD.13.2739} {\bibfield  {journal} {\bibinfo  {journal} {Phys.
  Rev. D}\ }\textbf {\bibinfo {volume} {13}},\ \bibinfo {pages} {2739--2761}
  (\bibinfo {year} {1976})}\BibitemShut {NoStop}%
\bibitem [{\citenamefont {Coleman}(1985)}]{Coleman:1985ki}%
  \BibitemOpen
  \bibfield  {author} {\bibinfo {author} {\bibfnamefont {Sidney~R.}\
  \bibnamefont {Coleman}},\ }\bibfield  {title} {\enquote {\bibinfo {title}
  {{Q-balls}},}\ }\href {\doibase 10.1016/0550-3213(86)90520-1} {\bibfield
  {journal} {\bibinfo  {journal} {Nucl. Phys. B}\ }\textbf {\bibinfo {volume}
  {262}},\ \bibinfo {pages} {263} (\bibinfo {year} {1985})},\ \bibinfo {note}
  {[Addendum: Nucl.Phys.B 269, 744 (1986)]}\BibitemShut {NoStop}%
\bibitem [{\citenamefont {Lee}\ and\ \citenamefont {Pang}(1992)}]{Lee:1991ax}%
  \BibitemOpen
  \bibfield  {author} {\bibinfo {author} {\bibfnamefont {T.~D.}\ \bibnamefont
  {Lee}}\ and\ \bibinfo {author} {\bibfnamefont {Y.}~\bibnamefont {Pang}},\
  }\bibfield  {title} {\enquote {\bibinfo {title} {{Nontopological
  solitons}},}\ }\href {\doibase 10.1016/0370-1573(92)90064-7} {\bibfield
  {journal} {\bibinfo  {journal} {Phys. Rept.}\ }\textbf {\bibinfo {volume}
  {221}},\ \bibinfo {pages} {251--350} (\bibinfo {year} {1992})}\BibitemShut
  {NoStop}%
\bibitem [{\citenamefont {Cohen}\ \emph {et~al.}(1986)\citenamefont {Cohen},
  \citenamefont {Coleman}, \citenamefont {Georgi},\ and\ \citenamefont
  {Manohar}}]{Cohen:1986ct}%
  \BibitemOpen
  \bibfield  {author} {\bibinfo {author} {\bibfnamefont {Andrew~G.}\
  \bibnamefont {Cohen}}, \bibinfo {author} {\bibfnamefont {Sidney~R.}\
  \bibnamefont {Coleman}}, \bibinfo {author} {\bibfnamefont {Howard}\
  \bibnamefont {Georgi}}, \ and\ \bibinfo {author} {\bibfnamefont {Aneesh}\
  \bibnamefont {Manohar}},\ }\bibfield  {title} {\enquote {\bibinfo {title}
  {{The Evaporation of $Q$ Balls}},}\ }\href {\doibase
  10.1016/0550-3213(86)90004-0} {\bibfield  {journal} {\bibinfo  {journal}
  {Nucl. Phys. B}\ }\textbf {\bibinfo {volume} {272}},\ \bibinfo {pages}
  {301--321} (\bibinfo {year} {1986})}\BibitemShut {NoStop}%
\bibitem [{\citenamefont {Kim}\ \emph {et~al.}(1993)\citenamefont {Kim},
  \citenamefont {Kim},\ and\ \citenamefont {Kim}}]{Kim:1992mm}%
  \BibitemOpen
  \bibfield  {author} {\bibinfo {author} {\bibfnamefont {Chan-ju}\ \bibnamefont
  {Kim}}, \bibinfo {author} {\bibfnamefont {Seyong}\ \bibnamefont {Kim}}, \
  and\ \bibinfo {author} {\bibfnamefont {Yoon-bai}\ \bibnamefont {Kim}},\
  }\bibfield  {title} {\enquote {\bibinfo {title} {{Global nontopological
  vortices}},}\ }\href {\doibase 10.1103/PhysRevD.47.5434} {\bibfield
  {journal} {\bibinfo  {journal} {Phys. Rev. D}\ }\textbf {\bibinfo {volume}
  {47}},\ \bibinfo {pages} {5434--5443} (\bibinfo {year} {1993})}\BibitemShut
  {NoStop}%
\bibitem [{\citenamefont {Kusenko}(1997)}]{Kusenko:1997ad}%
  \BibitemOpen
  \bibfield  {author} {\bibinfo {author} {\bibfnamefont {Alexander}\
  \bibnamefont {Kusenko}},\ }\bibfield  {title} {\enquote {\bibinfo {title}
  {{Small Q balls}},}\ }\href {\doibase 10.1016/S0370-2693(97)00582-0}
  {\bibfield  {journal} {\bibinfo  {journal} {Phys. Lett. B}\ }\textbf
  {\bibinfo {volume} {404}},\ \bibinfo {pages} {285} (\bibinfo {year}
  {1997})},\ \Eprint {http://arxiv.org/abs/hep-th/9704073}
  {arXiv:hep-th/9704073} \BibitemShut {NoStop}%
\bibitem [{\citenamefont {Dvali}\ \emph {et~al.}(1998)\citenamefont {Dvali},
  \citenamefont {Kusenko},\ and\ \citenamefont {Shaposhnikov}}]{Dvali:1997qv}%
  \BibitemOpen
  \bibfield  {author} {\bibinfo {author} {\bibfnamefont {G.~R.}\ \bibnamefont
  {Dvali}}, \bibinfo {author} {\bibfnamefont {Alexander}\ \bibnamefont
  {Kusenko}}, \ and\ \bibinfo {author} {\bibfnamefont {Mikhail~E.}\
  \bibnamefont {Shaposhnikov}},\ }\bibfield  {title} {\enquote {\bibinfo
  {title} {{New physics in a nutshell, or Q ball as a power plant}},}\ }\href
  {\doibase 10.1016/S0370-2693(97)01378-6} {\bibfield  {journal} {\bibinfo
  {journal} {Phys. Lett. B}\ }\textbf {\bibinfo {volume} {417}},\ \bibinfo
  {pages} {99--106} (\bibinfo {year} {1998})},\ \Eprint
  {http://arxiv.org/abs/hep-ph/9707423} {arXiv:hep-ph/9707423} \BibitemShut
  {NoStop}%
\bibitem [{\citenamefont {Kusenko}\ and\ \citenamefont
  {Shaposhnikov}(1998)}]{Kusenko:1997si}%
  \BibitemOpen
  \bibfield  {author} {\bibinfo {author} {\bibfnamefont {Alexander}\
  \bibnamefont {Kusenko}}\ and\ \bibinfo {author} {\bibfnamefont {Mikhail~E.}\
  \bibnamefont {Shaposhnikov}},\ }\bibfield  {title} {\enquote {\bibinfo
  {title} {{Supersymmetric Q balls as dark matter}},}\ }\href {\doibase
  10.1016/S0370-2693(97)01375-0} {\bibfield  {journal} {\bibinfo  {journal}
  {Phys. Lett. B}\ }\textbf {\bibinfo {volume} {418}},\ \bibinfo {pages}
  {46--54} (\bibinfo {year} {1998})},\ \Eprint
  {http://arxiv.org/abs/hep-ph/9709492} {arXiv:hep-ph/9709492} \BibitemShut
  {NoStop}%
\bibitem [{\citenamefont {Kusenko}\ \emph {et~al.}(1998)\citenamefont
  {Kusenko}, \citenamefont {Shaposhnikov},\ and\ \citenamefont
  {Tinyakov}}]{Kusenko:1997vi}%
  \BibitemOpen
  \bibfield  {author} {\bibinfo {author} {\bibfnamefont {Alexander}\
  \bibnamefont {Kusenko}}, \bibinfo {author} {\bibfnamefont {Mikhail~E.}\
  \bibnamefont {Shaposhnikov}}, \ and\ \bibinfo {author} {\bibfnamefont
  {P.~G.}\ \bibnamefont {Tinyakov}},\ }\bibfield  {title} {\enquote {\bibinfo
  {title} {{Sufficient conditions for the existence of Q balls in gauge
  theories}},}\ }\href {\doibase 10.1134/1.567658} {\bibfield  {journal}
  {\bibinfo  {journal} {Pisma Zh. Eksp. Teor. Fiz.}\ }\textbf {\bibinfo
  {volume} {67}},\ \bibinfo {pages} {229} (\bibinfo {year} {1998})},\ \Eprint
  {http://arxiv.org/abs/hep-th/9801041} {arXiv:hep-th/9801041} \BibitemShut
  {NoStop}%
\bibitem [{\citenamefont {Volkov}\ and\ \citenamefont
  {Wohnert}(2002)}]{Volkov:2002aj}%
  \BibitemOpen
  \bibfield  {author} {\bibinfo {author} {\bibfnamefont {Mikhail~S.}\
  \bibnamefont {Volkov}}\ and\ \bibinfo {author} {\bibfnamefont {Erik}\
  \bibnamefont {Wohnert}},\ }\bibfield  {title} {\enquote {\bibinfo {title}
  {{Spinning Q balls}},}\ }\href {\doibase 10.1103/PhysRevD.66.085003}
  {\bibfield  {journal} {\bibinfo  {journal} {Phys. Rev. D}\ }\textbf {\bibinfo
  {volume} {66}},\ \bibinfo {pages} {085003} (\bibinfo {year} {2002})},\
  \Eprint {http://arxiv.org/abs/hep-th/0205157} {arXiv:hep-th/0205157}
  \BibitemShut {NoStop}%
\bibitem [{\citenamefont {Clark}(2006)}]{Clark:2005zc}%
  \BibitemOpen
  \bibfield  {author} {\bibinfo {author} {\bibfnamefont {Stephen~S.}\
  \bibnamefont {Clark}},\ }\bibfield  {title} {\enquote {\bibinfo {title}
  {{Particle production from Q-balls}},}\ }\href {\doibase
  10.1016/j.nuclphysb.2006.08.019} {\bibfield  {journal} {\bibinfo  {journal}
  {Nucl. Phys. B}\ }\textbf {\bibinfo {volume} {756}},\ \bibinfo {pages}
  {38--70} (\bibinfo {year} {2006})},\ \Eprint
  {http://arxiv.org/abs/hep-ph/0510078} {arXiv:hep-ph/0510078} \BibitemShut
  {NoStop}%
\bibitem [{\citenamefont {'t~Hooft}(1974)}]{tHooft:1973alw}%
  \BibitemOpen
  \bibfield  {author} {\bibinfo {author} {\bibfnamefont {Gerard}\ \bibnamefont
  {'t~Hooft}},\ }\bibfield  {title} {\enquote {\bibinfo {title} {{A Planar
  Diagram Theory for Strong Interactions}},}\ }\href {\doibase
  10.1016/0550-3213(74)90154-0} {\bibfield  {journal} {\bibinfo  {journal}
  {Nucl. Phys. B}\ }\textbf {\bibinfo {volume} {72}},\ \bibinfo {pages} {461}
  (\bibinfo {year} {1974})}\BibitemShut {NoStop}%
\bibitem [{\citenamefont {Dvali}\ \emph
  {et~al.}(2015{\natexlab{b}})\citenamefont {Dvali}, \citenamefont {Gomez},
  \citenamefont {Isermann}, \citenamefont {L\"ust},\ and\ \citenamefont
  {Stieberger}}]{Dvali:2014ila}%
  \BibitemOpen
  \bibfield  {author} {\bibinfo {author} {\bibfnamefont {G.}~\bibnamefont
  {Dvali}}, \bibinfo {author} {\bibfnamefont {C.}~\bibnamefont {Gomez}},
  \bibinfo {author} {\bibfnamefont {R.~S.}\ \bibnamefont {Isermann}}, \bibinfo
  {author} {\bibfnamefont {D.}~\bibnamefont {L\"ust}}, \ and\ \bibinfo {author}
  {\bibfnamefont {S.}~\bibnamefont {Stieberger}},\ }\bibfield  {title}
  {\enquote {\bibinfo {title} {{Black hole formation and classicalization in
  ultra-Planckian 2\textrightarrow{}N scattering}},}\ }\href {\doibase
  10.1016/j.nuclphysb.2015.02.004} {\bibfield  {journal} {\bibinfo  {journal}
  {Nucl. Phys. B}\ }\textbf {\bibinfo {volume} {893}},\ \bibinfo {pages}
  {187--235} (\bibinfo {year} {2015}{\natexlab{b}})},\ \Eprint
  {http://arxiv.org/abs/1409.7405} {arXiv:1409.7405 [hep-th]} \BibitemShut
  {NoStop}%
\bibitem [{\citenamefont {Addazi}\ \emph {et~al.}(2017)\citenamefont {Addazi},
  \citenamefont {Bianchi},\ and\ \citenamefont {Veneziano}}]{Addazi:2016ksu}%
  \BibitemOpen
  \bibfield  {author} {\bibinfo {author} {\bibfnamefont {Andrea}\ \bibnamefont
  {Addazi}}, \bibinfo {author} {\bibfnamefont {Massimo}\ \bibnamefont
  {Bianchi}}, \ and\ \bibinfo {author} {\bibfnamefont {Gabriele}\ \bibnamefont
  {Veneziano}},\ }\bibfield  {title} {\enquote {\bibinfo {title} {{Glimpses of
  black hole formation/evaporation in highly inelastic, ultra-planckian string
  collisions}},}\ }\href {\doibase 10.1007/JHEP02(2017)111} {\bibfield
  {journal} {\bibinfo  {journal} {JHEP}\ }\textbf {\bibinfo {volume} {02}},\
  \bibinfo {pages} {111} (\bibinfo {year} {2017})},\ \Eprint
  {http://arxiv.org/abs/1611.03643} {arXiv:1611.03643 [hep-th]} \BibitemShut
  {NoStop}%
\bibitem [{\citenamefont {Brown}(1992)}]{Brown:1992ay}%
  \BibitemOpen
  \bibfield  {author} {\bibinfo {author} {\bibfnamefont {Lowell~S.}\
  \bibnamefont {Brown}},\ }\bibfield  {title} {\enquote {\bibinfo {title}
  {{Summing tree graphs at threshold}},}\ }\href {\doibase
  10.1103/PhysRevD.46.R4125} {\bibfield  {journal} {\bibinfo  {journal} {Phys.
  Rev. D}\ }\textbf {\bibinfo {volume} {46}},\ \bibinfo {pages} {R4125--R4127}
  (\bibinfo {year} {1992})},\ \Eprint {http://arxiv.org/abs/hep-ph/9209203}
  {arXiv:hep-ph/9209203} \BibitemShut {NoStop}%
\bibitem [{\citenamefont {Voloshin}(1992)}]{Voloshin:1992rr}%
  \BibitemOpen
  \bibfield  {author} {\bibinfo {author} {\bibfnamefont {M.~B.}\ \bibnamefont
  {Voloshin}},\ }\bibfield  {title} {\enquote {\bibinfo {title} {{Estimate of
  the onset of nonperturbative particle production at high-energy in a scalar
  theory}},}\ }\href {\doibase 10.1016/0370-2693(92)90901-F} {\bibfield
  {journal} {\bibinfo  {journal} {Phys. Lett. B}\ }\textbf {\bibinfo {volume}
  {293}},\ \bibinfo {pages} {389--394} (\bibinfo {year} {1992})}\BibitemShut
  {NoStop}%
\bibitem [{\citenamefont {Argyres}\ \emph {et~al.}(1993)\citenamefont
  {Argyres}, \citenamefont {Kleiss},\ and\ \citenamefont
  {Papadopoulos}}]{Argyres:1992np}%
  \BibitemOpen
  \bibfield  {author} {\bibinfo {author} {\bibfnamefont {E.~N.}\ \bibnamefont
  {Argyres}}, \bibinfo {author} {\bibfnamefont {Ronald H.~P.}\ \bibnamefont
  {Kleiss}}, \ and\ \bibinfo {author} {\bibfnamefont {Costas~G.}\ \bibnamefont
  {Papadopoulos}},\ }\bibfield  {title} {\enquote {\bibinfo {title} {{Amplitude
  estimates for multi - Higgs production at high-energies}},}\ }\href {\doibase
  10.1016/0550-3213(93)90140-K} {\bibfield  {journal} {\bibinfo  {journal}
  {Nucl. Phys. B}\ }\textbf {\bibinfo {volume} {391}},\ \bibinfo {pages}
  {42--56} (\bibinfo {year} {1993})}\BibitemShut {NoStop}%
\bibitem [{\citenamefont {Gorsky}\ and\ \citenamefont
  {Voloshin}(1993)}]{Gorsky:1993ix}%
  \BibitemOpen
  \bibfield  {author} {\bibinfo {author} {\bibfnamefont {A.~S.}\ \bibnamefont
  {Gorsky}}\ and\ \bibinfo {author} {\bibfnamefont {M.~B.}\ \bibnamefont
  {Voloshin}},\ }\bibfield  {title} {\enquote {\bibinfo {title}
  {{Nonperturbative production of multiboson states and quantum bubbles}},}\
  }\href {\doibase 10.1103/PhysRevD.48.3843} {\bibfield  {journal} {\bibinfo
  {journal} {Phys. Rev. D}\ }\textbf {\bibinfo {volume} {48}},\ \bibinfo
  {pages} {3843--3851} (\bibinfo {year} {1993})},\ \Eprint
  {http://arxiv.org/abs/hep-ph/9305219} {arXiv:hep-ph/9305219} \BibitemShut
  {NoStop}%
\bibitem [{\citenamefont {Libanov}\ \emph {et~al.}(1994)\citenamefont
  {Libanov}, \citenamefont {Rubakov}, \citenamefont {Son},\ and\ \citenamefont
  {Troitsky}}]{Libanov:1994ug}%
  \BibitemOpen
  \bibfield  {author} {\bibinfo {author} {\bibfnamefont {M.~V.}\ \bibnamefont
  {Libanov}}, \bibinfo {author} {\bibfnamefont {V.~A.}\ \bibnamefont
  {Rubakov}}, \bibinfo {author} {\bibfnamefont {D.~T.}\ \bibnamefont {Son}}, \
  and\ \bibinfo {author} {\bibfnamefont {Sergey~V.}\ \bibnamefont {Troitsky}},\
  }\bibfield  {title} {\enquote {\bibinfo {title} {{Exponentiation of
  multiparticle amplitudes in scalar theories}},}\ }\href {\doibase
  10.1103/PhysRevD.50.7553} {\bibfield  {journal} {\bibinfo  {journal} {Phys.
  Rev. D}\ }\textbf {\bibinfo {volume} {50}},\ \bibinfo {pages} {7553--7569}
  (\bibinfo {year} {1994})},\ \Eprint {http://arxiv.org/abs/hep-ph/9407381}
  {arXiv:hep-ph/9407381} \BibitemShut {NoStop}%
\bibitem [{\citenamefont {Libanov}\ \emph {et~al.}(1995)\citenamefont
  {Libanov}, \citenamefont {Son},\ and\ \citenamefont
  {Troitsky}}]{Libanov:1995gh}%
  \BibitemOpen
  \bibfield  {author} {\bibinfo {author} {\bibfnamefont {M.~V.}\ \bibnamefont
  {Libanov}}, \bibinfo {author} {\bibfnamefont {D.~T.}\ \bibnamefont {Son}}, \
  and\ \bibinfo {author} {\bibfnamefont {Sergey~V.}\ \bibnamefont {Troitsky}},\
  }\bibfield  {title} {\enquote {\bibinfo {title} {{Exponentiation of
  multiparticle amplitudes in scalar theories. 2. Universality of the
  exponent}},}\ }\href {\doibase 10.1103/PhysRevD.52.3679} {\bibfield
  {journal} {\bibinfo  {journal} {Phys. Rev. D}\ }\textbf {\bibinfo {volume}
  {52}},\ \bibinfo {pages} {3679--3687} (\bibinfo {year} {1995})},\ \Eprint
  {http://arxiv.org/abs/hep-ph/9503412} {arXiv:hep-ph/9503412} \BibitemShut
  {NoStop}%
\bibitem [{\citenamefont {Son}(1996)}]{Son:1995wz}%
  \BibitemOpen
  \bibfield  {author} {\bibinfo {author} {\bibfnamefont {D.~T.}\ \bibnamefont
  {Son}},\ }\bibfield  {title} {\enquote {\bibinfo {title} {{Semiclassical
  approach for multiparticle production in scalar theories}},}\ }\href
  {\doibase 10.1016/0550-3213(96)00386-0} {\bibfield  {journal} {\bibinfo
  {journal} {Nucl. Phys. B}\ }\textbf {\bibinfo {volume} {477}},\ \bibinfo
  {pages} {378--406} (\bibinfo {year} {1996})},\ \Eprint
  {http://arxiv.org/abs/hep-ph/9505338} {arXiv:hep-ph/9505338} \BibitemShut
  {NoStop}%
\bibitem [{\citenamefont {Dvali}(2018{\natexlab{b}})}]{Dvali:2018xoc}%
  \BibitemOpen
  \bibfield  {author} {\bibinfo {author} {\bibfnamefont {Gia}\ \bibnamefont
  {Dvali}},\ }\bibfield  {title} {\enquote {\bibinfo {title} {{Classicalization
  Clearly: Quantum Transition into States of Maximal Memory Storage
  Capacity}},}\ }\href@noop {} {\  (\bibinfo {year} {2018}{\natexlab{b}})},\
  \Eprint {http://arxiv.org/abs/1804.06154} {arXiv:1804.06154 [hep-th]}
  \BibitemShut {NoStop}%
\bibitem [{\citenamefont {Monin}(2018)}]{Monin:2018cbi}%
  \BibitemOpen
  \bibfield  {author} {\bibinfo {author} {\bibfnamefont {A.}~\bibnamefont
  {Monin}},\ }\bibfield  {title} {\enquote {\bibinfo {title} {{Inconsistencies
  of higgsplosion}},}\ }\href@noop {} {\  (\bibinfo {year} {2018})},\ \Eprint
  {http://arxiv.org/abs/1808.05810} {arXiv:1808.05810 [hep-th]} \BibitemShut
  {NoStop}%
\bibitem [{\citenamefont {Cornwall}(1990)}]{Cornwall:1990hh}%
  \BibitemOpen
  \bibfield  {author} {\bibinfo {author} {\bibfnamefont {John~M.}\ \bibnamefont
  {Cornwall}},\ }\bibfield  {title} {\enquote {\bibinfo {title} {{On the
  High-energy Behavior of Weakly Coupled Gauge Theories}},}\ }\href {\doibase
  10.1016/0370-2693(90)90850-6} {\bibfield  {journal} {\bibinfo  {journal}
  {Phys. Lett. B}\ }\textbf {\bibinfo {volume} {243}},\ \bibinfo {pages}
  {271--278} (\bibinfo {year} {1990})}\BibitemShut {NoStop}%
\bibitem [{\citenamefont {Goldberg}(1990)}]{Goldberg:1990qk}%
  \BibitemOpen
  \bibfield  {author} {\bibinfo {author} {\bibfnamefont {Haim}\ \bibnamefont
  {Goldberg}},\ }\bibfield  {title} {\enquote {\bibinfo {title} {{Breakdown of
  perturbation theory at tree level in theories with scalars}},}\ }\href
  {\doibase 10.1016/0370-2693(90)90628-J} {\bibfield  {journal} {\bibinfo
  {journal} {Phys. Lett. B}\ }\textbf {\bibinfo {volume} {246}},\ \bibinfo
  {pages} {445--450} (\bibinfo {year} {1990})}\BibitemShut {NoStop}%
\bibitem [{\citenamefont {Dvali}\ \emph
  {et~al.}(2015{\natexlab{c}})\citenamefont {Dvali}, \citenamefont {Gomez},
  \citenamefont {Gruending},\ and\ \citenamefont {Rug}}]{Dvali:2015jxa}%
  \BibitemOpen
  \bibfield  {author} {\bibinfo {author} {\bibfnamefont {Gia}\ \bibnamefont
  {Dvali}}, \bibinfo {author} {\bibfnamefont {Cesar}\ \bibnamefont {Gomez}},
  \bibinfo {author} {\bibfnamefont {Lukas}\ \bibnamefont {Gruending}}, \ and\
  \bibinfo {author} {\bibfnamefont {Tehseen}\ \bibnamefont {Rug}},\ }\bibfield
  {title} {\enquote {\bibinfo {title} {{Towards a Quantum Theory of
  Solitons}},}\ }\href {\doibase 10.1016/j.nuclphysb.2015.10.017} {\bibfield
  {journal} {\bibinfo  {journal} {Nucl. Phys. B}\ }\textbf {\bibinfo {volume}
  {901}},\ \bibinfo {pages} {338--353} (\bibinfo {year}
  {2015}{\natexlab{c}})},\ \Eprint {http://arxiv.org/abs/1508.03074}
  {arXiv:1508.03074 [hep-th]} \BibitemShut {NoStop}%
\bibitem [{\citenamefont {Dvali}(2016)}]{Dvali:2015aja}%
  \BibitemOpen
  \bibfield  {author} {\bibinfo {author} {\bibfnamefont {Gia}\ \bibnamefont
  {Dvali}},\ }\bibfield  {title} {\enquote {\bibinfo {title} {{Non-Thermal
  Corrections to Hawking Radiation Versus the Information Paradox}},}\ }\href
  {\doibase 10.1002/prop.201500096} {\bibfield  {journal} {\bibinfo  {journal}
  {Fortsch. Phys.}\ }\textbf {\bibinfo {volume} {64}},\ \bibinfo {pages}
  {106--108} (\bibinfo {year} {2016})},\ \Eprint
  {http://arxiv.org/abs/1509.04645} {arXiv:1509.04645 [hep-th]} \BibitemShut
  {NoStop}%
\bibitem [{\citenamefont {Page}(1993)}]{Page:1993wv}%
  \BibitemOpen
  \bibfield  {author} {\bibinfo {author} {\bibfnamefont {Don~N.}\ \bibnamefont
  {Page}},\ }\bibfield  {title} {\enquote {\bibinfo {title} {{Information in
  black hole radiation}},}\ }\href {\doibase 10.1103/PhysRevLett.71.3743}
  {\bibfield  {journal} {\bibinfo  {journal} {Phys. Rev. Lett.}\ }\textbf
  {\bibinfo {volume} {71}},\ \bibinfo {pages} {3743--3746} (\bibinfo {year}
  {1993})},\ \Eprint {http://arxiv.org/abs/hep-th/9306083}
  {arXiv:hep-th/9306083} \BibitemShut {NoStop}%
\bibitem [{\citenamefont {Dvali}(2018{\natexlab{c}})}]{Dvali:2017nis}%
  \BibitemOpen
  \bibfield  {author} {\bibinfo {author} {\bibfnamefont {Gia}\ \bibnamefont
  {Dvali}},\ }\bibfield  {title} {\enquote {\bibinfo {title} {{Area law
  microstate entropy from criticality and spherical symmetry}},}\ }\href
  {\doibase 10.1103/PhysRevD.97.105005} {\bibfield  {journal} {\bibinfo
  {journal} {Phys. Rev. D}\ }\textbf {\bibinfo {volume} {97}},\ \bibinfo
  {pages} {105005} (\bibinfo {year} {2018}{\natexlab{c}})},\ \Eprint
  {http://arxiv.org/abs/1712.02233} {arXiv:1712.02233 [hep-th]} \BibitemShut
  {NoStop}%
\end{thebibliography}%

\end{document}